\documentclass[PAPER, atlasdraft=false, UKenglish, coverpage=false, cernpreprint=true, orcidlogo, texmf, orcidlogo]{atlasdoc}
\usepackage[subcaption]{atlaspackage}
\usepackage{atlasbiblatex}

\usepackage{atlasphysics}
\graphicspath{{logos/}{figures/}}

\usepackage{ANA-TOPQ-2025-11-PAPER-defs}

\usepackage{lscape}
\usepackage{rotating}

\AtlasTitle{Observation of a cross-section enhancement near the $t\bar{t}$ production threshold in $\sqrt{s}=13$~TeV $pp$ collisions with the ATLAS detector}

\AtlasAbstract{%
A measurement of $t\bar{t}$ production is presented in the invariant-mass region near the pair production threshold, $m_{t\bar{t}} \sim 345$~GeV, in final states with two charged leptons and multiple jets. The measurement is based on $140\,\mathrm{fb}^{-1}$ of proton-proton collision data collected at $\sqrt{s} = 13$~TeV with the ATLAS detector at the Large Hadron Collider. The data are compared to two models of $t\bar{t}$ production:
a baseline model including only perturbative QCD predictions for the hard process at approximate next-to-next-to leading order accuracy in the strong coupling, and an extended model that, in addition, incorporates non-relativistic QCD simulations that also include the formation of colour-singlet quasi-bound-states near the $t\bar{t}$ threshold. The agreement between the data and the models is quantified via a profile-likelihood fit to the reconstructed $m_{t\bar{t}}$ distributions, in bins of two angular observables sensitive to spin-correlations in the $t\bar{t}$ system. An excess of events is observed over the baseline perturbative QCD prediction, with an observed significance over $8$ standard deviations. This excess is consistent with the formation of colour-singlet and spin-singlet $S$-wave quasi-bound $t\bar{t}$ states, as predicted by non-relativistic QCD, and corresponds to an observed cross-section of $9.3^{+1.4}_{-1.3}$~pb.
}

\AtlasRefCode{TOPQ-2025-11}

\PreprintIdNumber{CERN-EP-2026-002}

\AtlasJournalRef{Rep. Prog. Phys. 89 (2026) 057801}
\AtlasDOI{10.1088/1361-6633/ae60a0}

\AtlasCoverEgroupAnalysisTeam{atlas-topq-2025-11-analysis-team@cern.ch}

\hypersetup{pdftitle={ATLAS document},pdfauthor={The ATLAS Collaboration}}

\begin{document}

\maketitle

\tableofcontents

\clearpage

\section{Introduction}
%

The unprecedented centre-of-mass energies and luminosities of proton--proton collisions delivered by the Large Hadron Collider (LHC)~\cite{Evans:2008zzb} at CERN enable the study of the top quark, the heaviest known elementary particle, across a broad kinematic range.
The high top-quark production rate at the LHC, dominated by top--antitop production (\ttbar), and the excellent event reconstruction capabilities of the ATLAS detector allow for precision analyses across a variety of final states and different kinematic regimes.
This includes precision measurements and direct searches for new phenomena in distinct parts of the phase space, including previously inaccessible ones characterised by high Lorentz boosts of the top quarks.

In recent years, growing attention has turned to the opposite end of the kinematic spectrum where top quarks are produced with low velocities in the \ttbar centre-of-mass frame. In the case of top-quark pair (\ttbar) production, the dominant production mode for top quarks at the LHC, this regime corresponds to events in which the invariant mass of the system, \mttbar, is close to the production threshold, $\mttbar \approx 2 m_t \simeq 345$~\GeV.
In this kinematic regime, a large fraction of the \ttbar events are produced in a colour-singlet state and are maximally entangled~\cite{Kiyo:2008bv,Afik:2020onf}. By focusing exclusively on events with \mttbar\ near the production threshold, the first observation of quantum entanglement at the LHC became possible~\cite{TOPQ-2021-24,CMS-TOP-23-001}. Moreover, the \ttbar production cross-section in the threshold region is sensitive to the top-quark mass~\cite{Garzelli:2023rvx} and the top Yukawa coupling to the Higgs field, which can be constrained indirectly via measurements targeting this regime~\cite{CMS-TOP-17-004,TOPQ-2021-11}.

An intriguing property of this non-relativistic \ttbar system emerges when it is produced in a colour singlet state. In this case, the top quarks experience an attractive potential.
This phenomenon is well known for charm or bottom quarks, which form narrow quark--antiquark bound states, referred to as "charmonia" and "bottomia", respectively. These states appear as distinct narrow, resonant enhancements in the quark--antiquark production cross-section and typically decay via annihilation of the quark and antiquark.
By contrast, the large mass of the top quark and its correspondingly short lifetime strongly suppress the formation of \ttbar bound states because typically either the top or antitop quark decays via the weak interaction before a full bound state can be formed or quark--antiquark annihilation can occur~\cite{Bigi:1986jk}. Therefore, these \ttbar \textit{quasi-}bound-states, sometimes referred to as "toponia", do not appear in the \mttbar spectrum as individual sharp resonance peaks but as a local enhancement, broader than the hadronic resonances formed by the lighter quarks but with a width well below the current experimental resolution.
First hypothesised in the late 1980s~\cite{Fadin:1987wz, Fadin:1990wx, Sumino:1992ai}, well before the 1995 discovery of the top quark~\cite{D0:1995jca,CDF:1995wbb}, measurements of \ttbar quasi-bound-states would provide a stringent probe of quantum chromodynamics (QCD) in the non-relativistic regime (\nrQCD)~\cite{Caswell:1985ui,Bodwin:1994jh}. %

A key challenge of measurements and searches targeting the \ttbar production threshold lies in the accurate modelling of \nrQCD effects, as well as missing higher-order contributions in \pQCD, and off-shell effects in the top-quark decay. Calculations of the \ttbar invariant mass spectrum that incorporate \nrQCD effects were developed in the past~\cite{Hagiwara:2008df,Sumino:2010bv}, and now also include the resummation of soft and collinear gluon emissions near the threshold~\cite{Kiyo:2008bv, Garzelli:2024uhe}. They predict an enhancement of \ttbar production below the \ttbar threshold compared with next-to-leading-order (NLO) and next-to-next-to-leading-order (NNLO) \pQCD predictions due to the formation of quasi-bound-states.
The resummation of additional terms that scale in powers of $\alphas/v$~\cite{Nason:2025hix,Ju:2020otc} (where $v$ is the velocity of the top quarks) can also yield an enhancement of \ttbar production below the \ttbar threshold compared with current (N)NLO \pQCD predictions.
Additionally, off-shell effects and interference between \ttbar and the production of a single top quark with a $W$ boson ($tW$) at NLO in \pQCD are known to be sizeable close to threshold~\cite{Nason:2025hix}, especially with regard to \ttbar spin correlations. Predictions for a consistent NLO description of \ttbar and $tW$ production and decay, including quantum interference effects, were derived for both dileptonic~\cite{Jezo:2016ujg} and semileptonic~\cite{Jezo:2023rht} final states.

Tensions between data and current Monte Carlo (MC) models for \ttbar production were observed in several precision measurements and searches near the kinematic threshold by both the ATLAS~\cite{TOPQ-2018-17,TOPQ-2018-26,TOPQ-2021-24,EXOT-2020-25,TOPQ-2024-12} and CMS~\cite{CMS-TOP-20-006,CMS-TOP-23-001,CMS-TOP-23-007,CMS-HIG-22-013} Collaborations.
The MC models used in these measurements rely on matrix elements (ME) at NLO accuracy in \pQCD matched to a parton shower (PS) model.
In some cases, the NLO predictions were additionally reweighted to more accurate predictions at NNLO in \pQCD and NLO accuracy in the electroweak coupling. The CMS Collaboration has shown for two of these results~\cite{CMS-TOP-23-001,CMS-HIG-22-013} that the observed tension between the data and these models close to the production threshold can be alleviated by supplementing \pQCD predictions with a simplified model~\cite{Fuks:2021xje,Maltoni:2024tul,Aguilar-Saavedra:2024mnm} in which \ttbar quasi-bound-state effects are modelled as a narrow pseudo-scalar resonance, \toytoponium, with mass close to $2 \cdot \mtop$ and no interference with \pQCD \ttbar production.
Most recently, CMS reported an excess with a significance greater than five standard deviations near the \ttbar threshold compared with the \pQCD predictions, consistent with the production of a quasi-bound-state \toytoponium~\cite{CMS-TOP-24-007}.

In parallel, significant progress was made in providing a more complete MC model of contributions from \nrQCD at the \ttbar threshold, particularly concerning the inclusion of quasi-bound-states~\cite{Fuks:2024yjj,Fuks:2021xje,Fuks:2025wtq}.
In this approach, \nrQCD effects in colour-singlet \ttbar production, including the formation of quasi-bound-states, are described using a reweighting of the \ttbar production MEs for colour-singlet states through the non-relativistic QCD Green's function, which includes the resummation of terms proportional to $(\alphas/v)^k$.
The kinematics of the associated predictions for \ttbar quasi-bound-states using Green's Function Reweighted \ttbar, referred to as \toponium in the following, were shown to differ from those of the simplified pseudo-scalar model \toytoponium~\cite{Fuks:2025sxu}.
Adding these predictions to the standard \pQCD \ttbar MC provides a more complete prediction of the \ttbar threshold dynamics.

In this paper, a measurement of \ttbar production near the kinematic threshold
is presented, with a particular focus on \nrQCD effects, including the formation of \ttbar quasi-bound-states.
The analysis targets dileptonic \ttbar decays in events with two oppositely charged electrons or muons ($ee$, $e\mu$, $\mu\mu$). It is based on proton–proton ($pp$) collision data recorded with the ATLAS detector at a centre-of-mass energy of $\sqrt{s} = 13$~\TeV during LHC Run~2 (2015--2018), corresponding to an integrated luminosity of 140~fb$^{-1}$.
The measurement relies on \mttbar and two angular observables, \chel and \chan (Section~\ref{sec:selection}), evaluated in the \ttbar centre-of-mass frame, which are sensitive to \ttbar spin correlations. These observables were found to be powerful in selecting events with a spin-singlet contribution, which is the dominant contribution of the quasi-bound-state.
The inclusion of the \chel and \chan observables, together with the full \ttbar\ reconstruction, significantly enhances the sensitivity of this measurement to spin-singlet states compared with the search in Ref.~\cite{EXOT-2020-25} on the same data sample, which included dileptonic \ttbar\ decays without reconstructing the full \ttbar\ system and used a different set of spin-sensitive angular variables in both the dilepton and single-lepton channels.
A profile-likelihood fit is used to compare the level of agreement with data of two models of \ttbar production:
a baseline model including only PS-matched \pQCD predictions reweighted to NNLO accuracy in \alphas and NLO accuracy in the electroweak (EW) coupling (NNLO-QCD+NLO-EW); and an extended model that includes, in addition to the \pQCD predictions, state-of-the-art MC simulations of \nrQCD effects, including the formation of colour-singlet quasi-bound-states near the \ttbar production threshold~\cite{Fuks:2024yjj,Fuks:2021xje}.
The normalisation of the quasi-bound-state component in the extended model is a free-floating parameter in the fit.
Its fitted value is used to extract the total observed cross-section for the quasi-bound-state contribution.
%


%

%
\section{ATLAS detector}
%

%
%

%
\newcommand{\AtlasCoordFootnote}{%
ATLAS uses a right-handed coordinate system with its origin at the nominal interaction point (IP)
in the centre of the detector and the \(z\)-axis along the beam pipe.
The \(x\)-axis points from the IP to the centre of the LHC ring,
and the \(y\)-axis points upwards.
Cylindrical coordinates \((r,\phi)\) are used in the transverse plane,
\(\phi\) being the azimuthal angle around the \(z\)-axis.
The pseudorapidity is defined in terms of the polar angle \(\theta\) as \(\eta = -\ln \tan(\theta/2)\).
Angular distance is measured in units of \(\Delta R \equiv \sqrt{(\Delta\eta)^{2} + (\Delta\phi)^{2}}\).}

The ATLAS experiment~\cite{PERF-2007-01} at the LHC is a multipurpose particle detector
with a forward--backward symmetric cylindrical geometry and a near \(4\pi\) coverage in
solid angle.\footnote{\AtlasCoordFootnote}
It consists of an inner tracking detector surrounded by a thin superconducting solenoid
providing a \qty{2}{\tesla} axial magnetic field, electromagnetic and hadronic calorimeters, and a muon spectrometer.
The inner tracking detector covers the pseudorapidity range \(|\eta| < 2.5\).
It consists of silicon pixel, silicon microstrip, and transition radiation tracking detectors.
Lead/liquid-argon (LAr) sampling calorimeters provide electromagnetic (EM) energy measurements
with high granularity within the region \(|\eta|< 3.2\).
A steel/scintillator-tile hadronic calorimeter covers the central pseudorapidity range (\(|\eta| < 1.7\)).
The endcap and forward regions are instrumented with LAr calorimeters
for EM and hadronic energy measurements up to \(|\eta| = 4.9\).
The muon spectrometer surrounds the calorimeters and is based on
three large superconducting air-core toroidal magnets with eight coils each.
The field integral of the toroids ranges between \num{2.0} and \qty{6.0}{\tesla\metre}
across most of the detector.
The muon spectrometer includes a system of precision tracking chambers up to \(|\eta| = 2.7\) and fast detectors for triggering up to \(|\eta| = 2.4\).
The luminosity is measured mainly by the LUCID--2~\cite{LUCID2} detector which is located close to the beampipe.
A two-level trigger system was used to select events~\cite{TRIG-2016-01}.
The first-level trigger is implemented in hardware and used a subset of the detector information
to accept events at a rate close to \qty{100}{\kHz}.
This is followed by a software-based trigger that
reduced the accepted rate of complete events to \qty{1.25}{\kHz} on average
depending on the data-taking conditions.
A software suite~\cite{SOFT-2022-02} is used in data simulation, in the reconstruction
and analysis of real and simulated data, in detector operations, and in the trigger and data acquisition
systems of the experiment.


%

%
%

%
\section{Data and simulated event samples}
\label{sec:samples}

The data used in this measurement were collected with the ATLAS detector in $pp$ collisions at $\sqrt{s}=13$~\TeV during the years 2015 to 2018, corresponding to an integrated luminosity of 140~fb$^{-1}$~\cite{DAPR-2021-01}.
Candidate events were selected online using trigger algorithms requiring at least one muon~\cite{TRIG-2018-01} or one electron~\cite{TRIG-2018-05}, with transverse momentum (\pt) thresholds that were progressively raised during the data-taking period. The lowest un-prescaled electron trigger reached the efficiency plateau region for electrons with reconstructed $\pT > 25$~\GeV in 2015 and $\pT > 27$~\GeV in 2016–18. The corresponding thresholds for the muon trigger were 21~\GeV for 2015 and 27~\GeV thereafter.
These lowest un-prescaled lepton triggers included a lepton isolation requirement.
Only events recorded with all components of the ATLAS detector fully functional were selected.

Simulated event samples are used to model the different processes expected to contribute in the regions defined in the analysis.
For all MC samples, the ATLAS simulation infrastructure~\cite{SOFT-2010-01} was used.
All nominal samples and several alternative samples used to assess systematic uncertainties,
were produced with a detailed $\GEANT$~\cite{Agostinelli:2002hh} detector simulation, while a faster simulation was used for the remaining alternative background samples.
Pile-up effects are modelled by overlaying minimum-bias events simulated using the soft QCD processes of \PYTHIA[8.1]~\cite{Sjostrand:2007gs}
with the \NNPDF[2.3lo] set of parton distribution functions (PDFs)~\cite{Ball:2012cx} and the A3~\cite{ATL-PHYS-PUB-2016-017} tune.
The events are reweighted to match the pile-up conditions of each dataset corresponding to the years 2015--2018.
The same offline reconstruction methods used for data are applied to the simulated event samples.
Corrections are applied to match the selection efficiencies, energy and mass scales and resolutions of reconstructed simulated particles to those measured in data. %

The ME generators are interfaced with different models for the simulation of the PS, hadronisation and the underlying event.
Apart from events simulated with \SHERPA~\cite{Bothmann:2019yzt}, all MC samples use the \EVTGEN programme~\cite{Lange:2001uf} for the simulation of bottom and charm hadron decays.
Except for the pile-up effects discussed above, all events interfaced with \PYTHIA[8]~\cite{Sjostrand:2014zea,Bierlich:2022pfr} for the parton shower and hadronisation, the A14 tune~\cite{ATL-PHYS-PUB-2014-021} and the \NNPDF[2.3lo] set of PDFs are employed in the parton shower.
If not mentioned otherwise, the top-quark mass is set to 172.5~\GeV.

\subsection{Modelling of \texorpdfstring{$t\bar{t}$}{ttbar} production in \pQCD }
\label{subsec:ttbar_samples}

\subsubsection{Models based on \textsc{Powheg v2}\xspace\texttt{hvq}\xspace+\xspace\PYTHIA8\xspace (\phhvqpy)}
\label{sec:hvq}

The production of \ttbar events is modelled using the \POWHEGBOX[v2]~\cite{Re:2010bp,Frixione:2007nw,Nason:2004rx,Frixione:2007vw,Alioli:2010xd} generator with the \hvq model (\phhvq), which provides ME predictions for the production process $pp\rightarrow \ttbar$ at NLO accuracy in \pQCD.
For brevity, these samples are referred to in the text as \phhvqpy.
The \NNPDF[3.0nlo]~\cite{Ball:2014uwa} PDF set is used and the \hdamp parameter, which effectively regulates the high \pT radiation against which the \ttbar system recoils, is set to 1.5\mtop~\cite{ATL-PHYS-PUB-2016-020}.
The value of the strong coupling constant \alphas is taken at the reference value measured at the $Z$-boson mass scale, $\alphas(m_Z)=0.118$, and is consistently evolved to the relevant scales.
The renormalisation ($\mu_R$) and factorisation ($\mu_F$) scales are dynamic, using the functional form: $\sqrt{\mtop^{2} + p^2_{\mathrm{T}}(t)}$. The mass and transverse momentum of the top quark are taken before radiation.
The ME predictions from \phhvq are interfaced with \PYTHIA[8.2].
The value of the strong coupling constant in the final state shower of \PYTHIA ($\alpha_{s}^{\mathrm{FSR}}$) is set to 0.127. ME corrections that approximate NLO QCD are enabled in \PYTHIA for all emissions, compensating for the leading-order (LO) accuracy used in \phhvq to simulate the top-quark decay. The \pthard parameter, which affects the matching of the PS to the ME calculation, is set to 0.
The recoil for secondary and subsequent gluon emissions from the \bquark in the $t\to Wb$ vertex is assigned to the \bquark.\footnote{This corresponds to the setting \texttt{recoilToColor=on} in the \texttt{TimeShower} class in \PYTHIA.}
The early-resonance-decay parameter, which controls whether resonance decays can occur before or after colour reconnection (CR) happens in the simulation, is set such that resonance decays can only occur after CR\@.

Alternative \pQCD \phhvq \ttbar samples obtained with different generator choices and settings are used to estimate systematic uncertainties related to the modelling of this dominant contribution. The details can be found in Section~\ref{subsec:sys_tt}.

The \ttbar samples are normalised to the cross-section prediction at NNLO in QCD including the resummation of next-to-next-to-leading-logarithmic (NNLL) soft-gluon terms calculated using
\TOPpp[2.0]~\cite{Beneke:2011mq,Cacciari:2011hy,Baernreuther:2012ws,Czakon:2012zr,Czakon:2012pz,Czakon:2013goa,Czakon:2011xx}.
For $pp$ collisions at a centre-of-mass energy of \(\rts = \qty{13}{\TeV}\), this cross-section is calculated to be
\(\sigma(\ttbar)_{\text{NNLO+NNLL}} = 834^{+37}_{-43}\,\unit{\pb}\).
In this calculation, the PDF4LHC21 PDF set~\cite{Ball:2022oua} is used.
The uncertainty is obtained by summing in quadrature the uncertainties related to the PDF set, the uncertainties in the values of $\mu_R$, $\mu_F$, and \alphas, and the top-quark mass uncertainty of $\pm 1$~\GeV.
The \ttbar normalisation is a free-floating parameter in the fit, hence the central cross-section value and its uncertainty only affect the pre-fit predictions, not the final fitted rate.

Additionally, kinematic variables of the top-quark in the nominal \phhvqpy and all alternative pQCD \ttbar samples are corrected to more accurate differential predictions calculated at NNLO-QCD+NLO-EW accuracy.
The \MATRIX programme~\cite{Grazzini:2017mhc} is used to obtain the predictions at NNLO in \alphas,
while \HATHOR~\cite{Aliev:2010zk,Kuhn:2006vh,Kuhn:2013zoa,Kuhn:2005it}
is used to obtain the EW predictions at NLO.
In both programmes, $\mu_R$ and $\mu_F$ are set to the dynamic scale values of $\frac{1}{2}\sum_{i\in\left\{t,\bar{t}\right\}}\sqrt{\mtop^2+p_\mathrm{T,i}^2}$.
The predictions allow for a two-dimensional reweighting in $(\mttbar, \ctstar)$, where $\theta^*$ denotes the angle between the momentum of the top quark in the \ttbar centre-of-mass frame and the momentum of the reconstructed \ttbar system in the laboratory frame.
They are provided in the ranges $340 \leq \mttbar \leq 3500$~\GeV in bins of 10~\GeV and $-1 \leq \ctstar \leq +1$ in bins of 0.1.
Since both the NNLO-QCD and NLO-EW predictions assume that both top quarks are on-shell, no reweighting can be applied for events with $\mttbar < 340~\GeV$.

\subsubsection{Models based on \textsc{Powheg v2}\xspace\texttt{bb4l}\xspace+\xspace\PYTHIA8\xspace (\phbbfourlpy)}
\label{sec:bb4l}
Another \pQCD \ttbar sample was generated using the \texttt{bb4$\ell$} model implemented in \POWHEGBOXRES~\cite{Jezo:2015aia,Jezo:2016ujg,Jezo:2023rht} (\phbbfourl).
These samples are referred to in the text as \phbbfourlpy.
This model includes off-shell and non-resonant contributions as well as exact spin correlations at NLO accuracy and simulates the inclusive production of $b\bar{b}\ell^+\ell^-\nu\bar{\nu}$ final states.
In this way interference effects between \ttbar and $tW$ are taken into account.
It also uses the new inverse-width correction described in Ref.~\cite{Jezo:2023rht} to correct for spurious width effects.
The same settings as those used in the nominal \ttbar sample are adopted for the renormalisation and factorisation scales, top-quark mass, PDF set, and value of \alphas. The $bb4\ell$ predictions are interfaced with \PYTHIA[8.3] using the A14 tune and the same PDF set as in the nominal sample. The $bb4\ell$ sample uses a recoil scheme in which the top-quark participates in the recoil to QCD radiation from $b$-quarks\footnote{This corresponds to the setting \texttt{recoilStrategyRF=3} in the \texttt{TimeShower} class in \PYTHIA[8.3].} and the $p_{\textrm T}^\textrm{maxMatch}$ and the $p_{\textrm T}^\textrm{def}$ parameters\footnote{Setting $p_{\textrm T}^\textrm{maxMatch}$ to 2 allows emissions up to the kinematic limit, the setting for $p_{\textrm T}^\textrm{def}$ enables the use of the \PYTHIA\ $p_{\textrm T}$ definition.} are both set to 2.
This sample is used to evaluate a systematic uncertainty related to the modelling of top-quark decays and off-shell effects (Section~\ref{sec:systematics}). Additionally, it is used to provide an alternative baseline \ttbar \pQCD prediction, replacing the nominal \phhvqpy setup, which is compared with the data in Appendix~\ref{sec:app_bb4l}. The detailed settings used to generate the sample are specified in Ref.~\cite{ATL-PHYS-PUB-2025-036}.

The \phbbfourlpy sample is normalised to the sum of the calculated higher-order cross-sections
for \ttbar (Section~\ref{sec:hvq}) and $tW$ (Section~\ref{subsec:background_samples}) production.
An alternative cross-section calculation for \bbfourl at approximate NNLO-QCD accuracy using the double-pole approximation has recently become available~\cite{Buonocore:2025fqs}. The cross-section value obtained from this calculation is around 4.5\% below the sum of the calculated higher-order cross-sections
for \ttbar and $tW$ and compatible with it within one standard deviation of the relevant uncertainties.
Kinematic distributions are not reweighted to higher-order calculations, since isolating \ttbar events within the \phbbfourlpy sample, which also includes $tW$ contributions and their interference, would affect the overall modelling accuracy.

%
%

%

%
%
%
%
%
%
%
%
%
%
%
%
%
%
%


\subsection{Modelling of \texorpdfstring{$t\bar{t}$}{ttbar} quasi-bound-state effects}
\label{subsec:signal_samples}

At the LHC, the dominant \nrQCD contribution to the \ttbar cross-section near the production threshold arises from the production of colour-singlet \ttbar pairs via gluon--gluon fusion ($gg \rightarrow \ttbar$), which experience an attractive QCD potential, leading to the formation of quasi-bound-states.
Although top--antitop quark pairs produced in a colour-octet configuration experience a repulsive QCD potential, gluon exchange between the pair still leads to an effect on the cross-section, though it is subleading compared with the enhancement arising from colour-singlet states.

In the present analysis, \nrQCD effects close to the production threshold are modelled following the approach described in Ref.~\cite{Fuks:2024yjj}.
In this framework, the formation of colour-singlet states is described using the Green’s function of non-relativistic QCD in the Coulomb gauge, which determines the momentum distribution of top quarks under the influence of the QCD potential.
The Green's function is computed numerically by solving the Lippmann-Schwinger equation with input from the Fourier transform of the QCD potential~\cite{Fuks:2021xje,Jezabek:1992np,Hagiwara:2016rdv,Hagiwara:2017ban}.
The result is used to reweight the MEs obtained at LO accuracy in \pQCD for the production of a (potentially off-shell) colour-singlet \ttbar pair in an $S$-wave configuration near the \ttbar threshold, to simulate a sample of quasi-bound-state \ttbar events.
The model includes the full spectrum of $^1S_0$ mass states around the \ttbar production threshold. It accounts only for gluon-initiated processes, which can produce $^1S_0$ states, while subdominant quark–antiquark–initiated contributions to $^3S_1$ states are not included.

The full production and decay into a six-body final state is simulated using \MADGRAPH[3.4.2]~\cite{Alwall:2014hca} with the \NNPDF[3.0nlo] PDF set and interfaced with \PYTHIA[8.3].
For consistency with the calculations in Ref.~\cite{Fuks:2024yjj}, which are used as input to the simulation of this sample, the top-quark mass and width are set to $173~\GeV$ and $1.49~\GeV$, respectively.
The value of the strong coupling constant in these calculations is set to $\alphas(m_Z)=0.12$.
As recommended in Ref.~\cite{Fuks:2024yjj}, the Green's function reweighting is applied only to events with $\mttbar < 350$~\GeV and a top-quark momentum magnitude $p^*$ in the \ttbar rest frame of less than 50~\GeV, evaluated at the parton level. Such kinematic requirements are not applied to any other MC sample used in this measurement.
The overlap between the \toponium sample and the \pQCD \ttbar samples described in Sections~\ref{sec:hvq} and~\ref{sec:bb4l} is assumed to be minimal, hence the double-counting corrections suggested in Ref.~\cite{Fuks:2024yjj} are not applied.%

The model described above provides a more complete description of \ttbar quasi-bound-state formation than the simplified models introduced in Refs.~\cite{Fuks:2021xje, Maltoni:2024tul} that are used in Refs.~\cite{CMS-TOP-23-001,CMS-TOP-23-007,CMS-TOP-24-007} and compared in Ref.~\cite{Fuks:2025sxu}.
In these simplified models, \ttbar quasi-bound-states are represented either as a pseudo-scalar spin-singlet resonance or as a combination of scalar and pseudo-scalar resonances.
These simplified models are motivated by the observation that the dominant $S$-wave contribution to $gg \to \ttbar$ production near the threshold arises mostly from the lightest pseudo-scalar $^1S_0^{[1]}$ configuration~\cite{Kiyo:2008bv, Sumino:2010bv}.
The contributions from states with higher total angular momentum $J$ and colour-octet states are expected to be sub-dominant and are neglected.
In contrast, the nominal \ttbar quasi-bound-state model, \toponium, used in this analysis includes the full set of $S$-wave colour-singlet contributions from gluon-initiated processes,
incorporating both bound-state and scattering-state effects.

The sample is normalised to a cross-section of 6.43~pb\footnote{The value of the cross-section depends significantly on the chosen \mttbar integration range.} (including all decay channels), obtained from analytical calculations~\cite{Fuks:2021xje}.
No uncertainty estimate is currently available for this value.
This calculated cross-section is larger than the cross-section of 5.60~pb obtained directly from the MC simulation of this model that does not include contributions from $P$-wave and colour-octet states.
A branching ratio of 10.82\% per lepton flavour is assumed in the normalisation to the calculated inclusive cross-section.

To facilitate a direct comparison with the CMS analysis of Ref.~\cite{CMS-TOP-24-007}, an alternative \ttbar quasi-bound-state sample was simulated using the simplified model of Ref.~\cite{Fuks:2021xje,Maltoni:2024tul} with a pseudo-scalar resonance \toytoponium. The sample was produced in \MADGRAPH[3.5.5] with the \NNPDF[3.0nlo] PDF set. The \toytoponium mass was set to 343~\GeV, in line with the expectation that the mass of the \ttbar quasi-bound ground state is twice that of the top quark minus a binding energy of about 2~\GeV~\cite{Sumino:2010bv, Fuks:2021xje}. Its total width was set to 2.8~\GeV, i.e. roughly twice that predicted for the top quark~\cite{Fadin:1990wx}. No kinematic requirements on \mttbar or $p^*$ were applied.
The full production and decay into the four-body $WbWb$ final state was simulated. The PS and hadronisation were modelled with \PYTHIA[8.3]. %
As in the nominal \toponium model, the sample cross-section is set to 6.43~pb.

Recently, another MC model of quasi-bound-state formation near the production threshold has become available in \PYTHIA[8.316]~\cite{Sjostrand:2025qez}. It provides \nrQCD corrections close to threshold for both colour-singlet and -octet states, derived with the formalism developed in Ref.~\cite{Fadin:1990wx}. It does not, however, account for spin correlations between the top quark and antiquark, and it is not used in this paper. %
%


\subsection{Modelling of other Standard Model processes}
\label{subsec:background_samples}

Single-top \(tW\) associated production is modelled at NLO in \pQCD using the \POWHEGBOX[v2] generator in the five-flavour scheme with the \NNPDF[3.0nlo] PDF.
The renormalisation and factorisation scales are dynamically set to $\HT/2$, where \HT is defined as the scalar sum of
the \pt and invariant masses of the top quark and $W$ boson~\cite{ATL-PHYS-PUB-2021-042}.
The diagram-removal scheme~\cite{Frixione:2008yi} is employed to handle the interference with \ttbar production~\cite{ATL-PHYS-PUB-2016-020}.
The events are interfaced with \PYTHIA[8.2] for the PS and hadronisation.
The inclusive cross-section is corrected to the theory prediction of $\sigma(tW)=79.3^{+2.9}_{-2.8}$\,pb, computed at NLO in QCD with the addition of third-order corrections of soft-gluon emissions by resummation of NNLL terms~\cite{Kidonakis:2021vob}.
The top-quark mass is set to 172.5~\GeV and the PDF4LHC21 PDF set is used.
The quoted uncertainty includes variations in the scales $\mu_R$, $\mu_F$, and the PDFs.

Single-top \(t\)-channel production is modelled using \POWHEGBOX[v2] to provide ME predictions at NLO in \pQCD using the four-flavour
scheme with the corresponding \NNPDF[3.0nlo] PDF set.
The functional form of $\mu_R$ and $\mu_F$ is set to
\(4\sqrt{m_b^2 + \pTX[2][b]}\) following the recommendation of Ref.~\cite{Frederix:2012dh}. Top quarks are decayed at LO using \MADSPIN~\cite{Frixione:2007zp,Artoisenet:2012st}.
The events are interfaced with \PYTHIA[8.2] for the PS and hadronisation.
The cross-section is applied according to the calculation in Ref.~\cite{ATL-PHYS-PUB-2025-035}.

Single-top \(s\)-channel production are modelled using \POWHEGBOX[v2] to provide MEs at NLO in \pQCD in the five-flavour scheme with the \NNPDF[3.0nlo] PDF set, while $\mu_R$ and $\mu_F$ are set to $\mtop$. The events are interfaced with \PYTHIA[8.2].
The inclusive cross-section is corrected to the theory predictions calculated to approximate NNLO in QCD~\cite{Liu:2018gxa}.

The production of a $Z$ or $W$ boson in association with jets (\(V+\)jets) are simulated with the \SHERPA[2.2.11] or \SHERPA[2.2.14]~\cite{Bothmann:2019yzt}
generator using NLO MEs for up to two partons, and LO MEs for up to five partons calculated in the five-flavour scheme with the \textsc{Comix}~\cite{Gleisberg:2008fv}
and \OPENLOOPS~\cite{Buccioni:2019sur,Cascioli:2011va,Denner:2016kdg} libraries. These predictions are matched with the \SHERPA PS~\cite{Schumann:2007mg} using the \MEPSatNLO prescription~\cite{Hoeche:2011fd,Hoeche:2012yf,Catani:2001cc,Hoeche:2009rj}
and the set of tuned parameters recommended for \SHERPA.
The \NNPDF[3.0nnlo] PDF set is used and the samples are normalised to the cross-section calculated at NNLO in \pQCD~\cite{Anastasiou:2003ds}.
The $Z$+jets samples include contributions from virtual photon exchange and their interference with processes involving intermediate $Z$ bosons.

Diboson (\(VV\)) productions are simulated with
\SHERPA[2.2.1] or 2.2.2 depending on the process,
including off-shell effects and Higgs-boson contributions where appropriate.
Fully leptonic final states and semileptonic final states, where one boson
decays leptonically and the other hadronically, are generated at NLO accuracy in \pQCD for up to one additional parton
and at LO accuracy in \pQCD for up to three additional parton emissions. Samples for the loop-induced processes \(gg \to VV\) are generated at LO accuracy for up to one additional parton emission for both fully leptonic and semileptonic final states. The ME calculations are matched and merged with the \SHERPA PS based on Catani--Seymour dipole factorisation~\cite{Gleisberg:2008fv,Schumann:2007mg} using the MEPS@NLO prescription~\cite{Hoeche:2011fd,Hoeche:2012yf,Catani:2001cc,Hoeche:2009rj}.
The virtual QCD corrections are taken from the
\OPENLOOPS library~\cite{Buccioni:2019sur,Cascioli:2011va,Denner:2016kdg}. The
\NNPDF[3.0nnlo] set of PDFs is used, along with the dedicated set of tuned parton-shower parameters recommended for \SHERPA.

Other small backgrounds include the production of \ttH events, modelled at NLO in \pQCD using
\POWHEGBOX[v2] with the \NNPDF[3.0nlo] PDF set,
interfaced with \PYTHIA[8.2] for the PS and hadronisation.
The samples are normalised to the calculated total \ttH cross-section
at NLO QCD+EW accuracy~\cite{deFlorian:2016spz}.
The production of \ttV\ events is modelled at NLO in \pQCD using the
\MGNLO[2.3.3]~\cite{Alwall:2014hca}
generator with the \NNPDF[3.0nlo] PDF set,
interfaced with \PYTHIA[8.2].
The samples are normalised to the calculated cross-sections for \ttbar{}+$Z$ and \ttbar{}+$W$ production, which are available at NLO in \pQCD and NLO in the EW coupling~\cite{deFlorian:2016spz}.
The production of $tWZ$ events is modelled at NLO in \pQCD using \MGNLO[2.3.3] with the \NNPDF[3.0nlo] PDF set, interfaced with \PYTHIA[8.2].
Finally, $tZq$ events were generated at NLO in \pQCD using \MGNLO[2.9.16] in the 4-flavour scheme and interfaced with \PYTHIA[8.3].



%

%
\section{Object reconstruction}
\label{sec:event}

Common event-quality criteria and object definitions are applied to all events considered for further analysis,
including standard data-quality requirements to select data events
with the detector in good operating condition~\cite{DAPR-2018-01}.
Additional event selection criteria that are specific to the objects and kinematic variables of interest in dileptonic \ttbar events are applied as described in Section~\ref{sec:selection}.

Events are required to have at least one reconstructed $pp$ interaction vertex
with a minimum of two associated tracks with transverse momenta $\pT > 0.5$~\GeV.
The primary vertex is defined as the vertex with the highest sum of squared transverse momenta of associated tracks~\cite{ATL-PHYS-PUB-2015-026}.

Muon candidates are reconstructed from matching tracks in the inner detector with tracks or track segments in the muon spectrometer, refined through a global fit that uses the hits from both subdetectors.
They must have $\pt>10$~\GeV\ and $|\eta|<2.5$, and satisfy a set of ``Medium'' identification criteria~\cite{MUON-2018-03}.
The longitudinal track impact parameter $z_0$ is required to satisfy
$|z_0\sin\theta| <0.5$~mm and $|d_0/\sigma(d_0)|<3$ is required for the transverse impact parameter $d_0$ and its uncertainty $\sigma(d_0)$.
Muons are required to be isolated~\cite{MUON-2018-03}; that is, the sum of the transverse momenta of the tracks within a variable-radius cone around the muon direction, excluding the muon track, must be less than 6\% of the transverse momentum of the muon.
The track isolation cone size is given by the minimum of $R = 10\;\mathrm{GeV}/\pt^{\mu}$ and $R=0.3$.
Electron candidates are reconstructed from energy deposits in the electromagnetic calorimeter matched to a charged-particle track in the inner detector~\cite{EGAM-2018-01}.
The track is required to be matched to the primary vertex by imposing the requirement $|z_0 \sin\theta| < 0.5$~mm and $|d_0/\sigma(d_0)|<5$.
Electron candidates are required to be within $|\eta_{\textrm{cluster}}|<2.47$, excluding the transition region between the barrel and endcap calorimeters ($1.37 < |\eta_{\textrm{cluster}}| < 1.52$).
They are further required to satisfy $\pt>10$~\GeV\ and the ``TightLH'' likelihood identification criteria.
The same variable-cone isolation requirement as for muons is imposed on electrons, with the exception that the maximum cone radius is set to 0.2.

Jets are reconstructed from particle-flow objects~\cite{PERF-2015-09,JETM-2018-05} using the anti-$k_t$ algorithm~\cite{Cacciari:2008gp,Fastjet} with a radius parameter $R=0.4$.
The energy scale of jets is calibrated using both simulation and data~\cite{JETM-2018-05}. Contributions from pile-up are subtracted using the jet-area method~\cite{Cacciari:2007fd}. Corrections are also applied to the simulation to bring the jet energy resolution in simulation into agreement with the jet energy resolution measured in data~\cite{JETM-2018-05}.
To suppress jets arising from pile-up, a jet-vertex-tagging (JVT) technique using a multivariate likelihood~\cite{PERF-2014-03} is applied to jets with $\pt < 60$~\GeV and $|\eta|<2.4$, ensuring that selected jets are matched to the primary vertex.
The jets are selected by requiring $\pt>25$~\GeV\ and $|\eta|<2.5$.

Jets likely to contain decays of a $b$-hadron (henceforth called $b$-tagged) are identified using the ``DL1r'' algorithm~\cite{FTAG-2019-07}.
This algorithm is based on a multivariate classification technique with a deep neural network combining information from the impact parameters of tracks
and topological properties of secondary and tertiary decay vertices reconstructed from the tracks associated with the jet.
A working point (WP) corresponding to a $b$-tagging efficiency of 70~\% is used. It provides rejection factors of approximately 380 for light-flavour jets and 10 for $c$-jets~\cite{FTAG-2019-07}.
Efficiency and rejection correction factors are applied to the simulated jets, depending on whether they are $b$-tagged~\cite{FTAG-2018-01} or mistagged~\cite{FTAG-2020-08,FTAG-2019-02}.
The correction for $b$-jets is derived from \ttbar events with final states containing two leptons, and the corrections are consistent with unity within uncertainties at the level of a few percent over most of the jet \pt range.

The reconstruction of the same energy deposits as multiple objects is resolved using an overlap removal procedure.
First, the closest selected jet within $\Delta R_y < 0.2$ of an electron is removed.\footnote{The quantity $\Delta R_y$ is defined in analogy to the angular separation $\Delta R$ but using the rapidity $y$ instead of the pseudorapidity $\eta$.}
Next, if an electron is found to be close to a jet ($\Delta R_y(\textrm{electron}, \textrm{jet}) < 0.4$), the electron is removed.
Then jets that have fewer than three matched tracks are removed if they are within $\Delta R_y <$~0.2 of a muon candidate.
Finally, muons that are within $0.2 < \Delta R_y < 0.4$ of a remaining jet are removed to suppress non-prompt muons.

The missing transverse momentum, \MET, is defined as the magnitude of the negative vectorial sum of the transverse momenta, \ptmiss, of all physics objects defined above.
Tracks that are not associated with these physics objects but originate from the primary vertex are taken into account as a soft term~\cite{JETM-2020-03}.


%

%
\section{Event selection and categorisation}
\label{sec:selection}

Events with a detector signature consistent with that expected to arise from dileptonic \ttbar events are selected by first imposing a set of loose preselection requirements.
They are required to have fired one of the single-electron or single-muon triggers described in Section~\ref{sec:samples}.
Candidate events are further required to have exactly two charged leptons (electrons or muons) and at least two reconstructed jets.
Depending on the lepton flavour, events are referred to as $ee$, $\mu\mu$, or $e\mu$ events.
At least one of these leptons is required to have $\pt>25/27/28$~\GeV for data taken in 2015, 2016 and 2017+2018, respectively, to ensure that the single-lepton triggers are fully efficient.
Additionally, this lepton is required to match within $\Delta R < 0.15$ the lepton with the same flavour reconstructed by the trigger algorithm.
Additionally, at least one of the jets is required to be $b$-tagged.
The invariant mass of the reconstructed \ttbar system (Section~\ref{sec:reco_2l}) must satisfy $\mttbar\leq 500~\GeV$ to prevent potential mis-modellings at high \mttbar values from impacting the measurement.

The events are then further categorised into a set of signal regions (SRs) and two sets of control regions (CRs), used to estimate the background from $Z$+jets events (CR-Z) and from non-prompt and fake-lepton events (CR-Fakes).

Events in the SRs are required to contain two leptons with opposite electric charge, i.e. $e^+e^-$, $\mu^+\mu^-$, $e^+\mu^-$ or $e^-\mu^+$.
Additionally, for $e^+e^-$ and $\mu^+\mu^-$ events, the dilepton invariant mass, \mll, is required to be greater than 15~\GeV, not within the range 81--101~\GeV around the $Z$-boson mass and the event is required to have $\met \geq 60$~\GeV, to reduce $Z$+jets events.

The CR-Z region comprises opposite-charge $ee$ and $\mu\mu$ events compatible with the production of an on-shell $Z$ boson.
This is achieved by inverting the $Z$-boson mass-window requirement of the SRs.
The same \met requirement as for the SRs is applied here.
Finally, three CR-Fakes regions are constructed to contain same-charge $ee$, $\mu\mu$, and $e\mu$ events, respectively.
For $ee$ and $\mu\mu$ events, the dilepton invariant mass is required to be outside the range 81--101~\GeV and additionally required to be greater than 15~\GeV.
No \met requirement is applied for the CR-Fakes regions.
The contributions from quasi-bound-state formation to the overall \ttbar contributions in the different CRs were found to be negligible. The event selection requirements are summarised in Table~\ref{tab:selection_2l}.

\begin{table}[h]
\centering
\caption{Summary of event selection criteria for the SRs and CRs in the analysis. The letter $\ell$ refers to $e$ and $\mu$, and OSSF refers to a lepton pair with opposite charge and the same flavour.}
\begin{tabular}{c|c|c}
\toprule
\textbf{SRs} & \textbf{CR-Z} & \textbf{CR-Fakes} \\
\midrule
\multicolumn{3}{c}{$=2\ell$ with $\pt(\ell) \geq 10~\GeV$} \\
\multicolumn{3}{c}{$\geq 1$ trigger-matched lepton with $\pt \geq 25/27/28~\GeV$} \\
\multicolumn{3}{c}{$\geq 2$ jets with $\pt \geq 25~\GeV$} \\
\multicolumn{3}{c}{$\geq 1$ $b$-tagged jet (70\% efficiency WP)} \\
\multicolumn{3}{c}{$m_{\ell\ell} \geq 15~\GeV$ for $ee$ and $\mu\mu$ events} \\
\multicolumn{3}{c}{$m_{t\bar{t}} \leq 500~\GeV$} \\
\midrule
\multicolumn{2}{c|}{$\met \geq 60~\GeV$ for OSSF events} & --- \\
\midrule
$\ell^\pm\ell^{'\mp}$ & $e^\pm e^\mp/\mu^\pm\mu^\mp$ & $\ell^\pm\ell^{'\pm}$ \\
$|m_{\ell\ell} - m_Z| \geq 10~\GeV$ & $|m_{\ell\ell} - m_Z| \leq 10~\GeV$ & $|m_{\ell\ell} - m_Z| \geq 10~\GeV$ \\
\bottomrule
\end{tabular}
\label{tab:selection_2l}
\end{table}

\begin{figure}[htbp]
\centering
\begin{subfigure}{0.45\textwidth}
\centering
\includegraphics[width=0.98\textwidth]{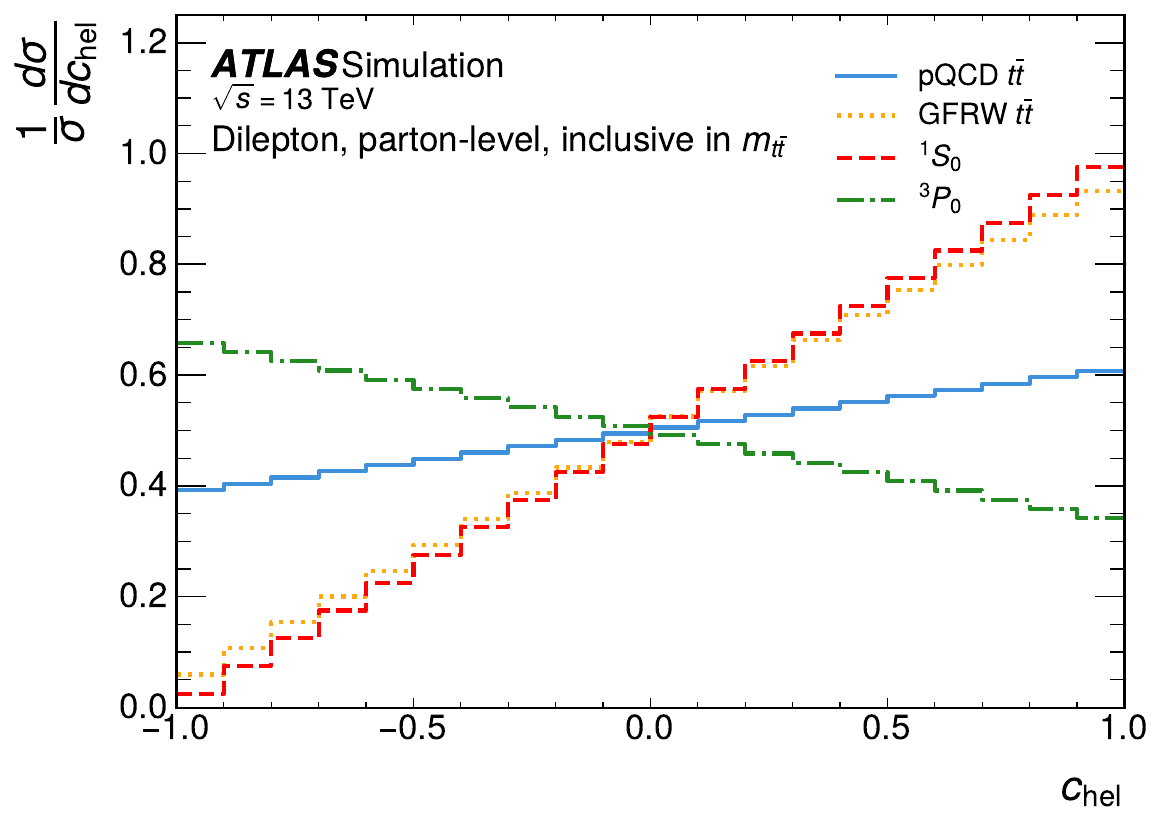}
\caption{}
\end{subfigure}
\begin{subfigure}{0.45\textwidth}
\centering
\includegraphics[width=0.98\textwidth]{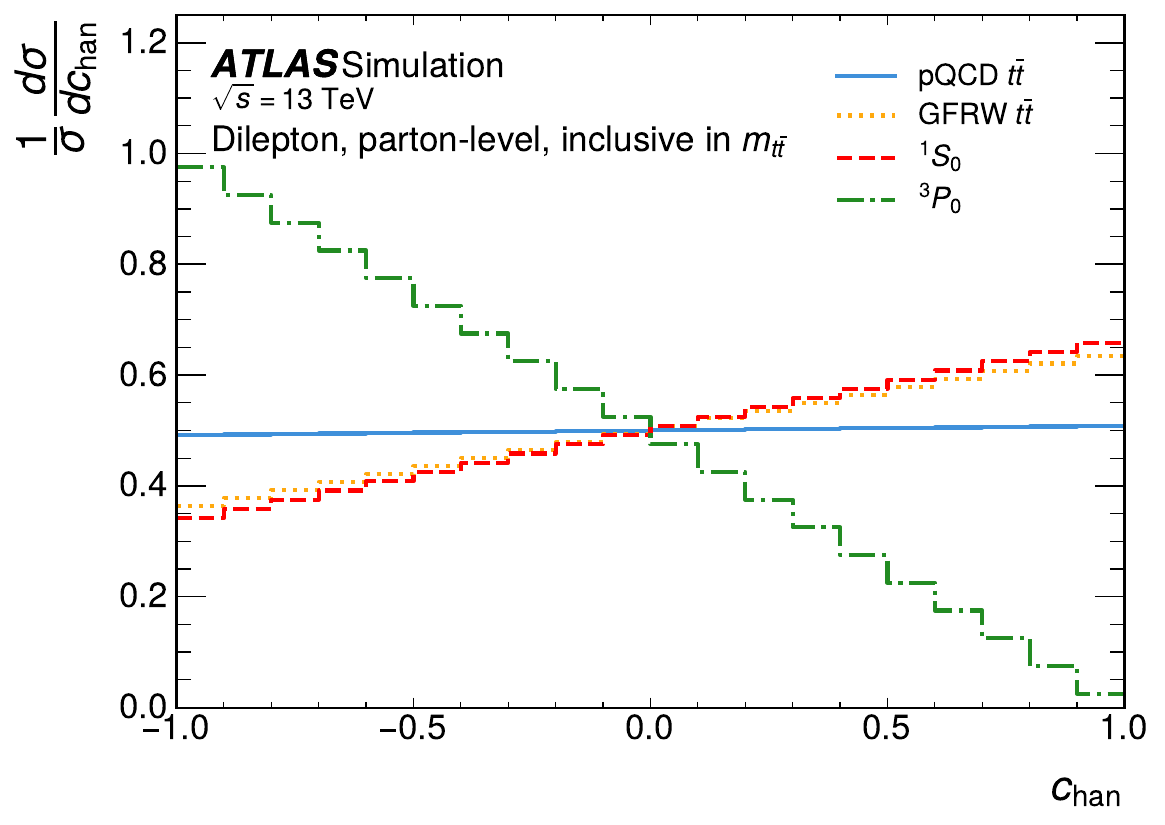}
\caption{}
\end{subfigure}
\caption{Parton-level distributions of (a) \chel and (b) \chan for \pQCD \ttbar and \toponium production, emphasizing their different shapes. The distributions for a pure spin-singlet state ($^1S_0$) and a pure spin-triplet state ($^3P_0$), obtained analytically, are shown as well. The \toponium sample includes the full spectrum of $^1S_0$ states, and therefore has a shape similar to the analytical $^1S_0$ distribution.
}
\label{fig:sel:chelchan}
\end{figure}

Events in the SR are further categorised into nine angular regions based on the reconstructed values of the observables \chel and \chan. Details on the reconstruction of the \ttbar system can be found in Section~\ref{sec:reco_2l}.
The observable \chel is the cosine of the angle between the charged-lepton momenta after Lorentz boosting the charged-lepton four-momenta first into the \ttbar centre-of-mass frame, and then separately into their parent top-quark and antitop-quark’s rest frames.
It is the same angle used in the ATLAS and CMS \ttbar entanglement analyses in \ttbar dileptonic final states~\cite{TOPQ-2021-24,CMS-TOP-23-001}.
A linear distribution is expected, with a slope proportional to the trace of the spin correlation matrix of the \ttbar pair.
For a spin-singlet, the spins are maximally anti-correlated, and therefore the diagonal spin-correlations are $-1$.
Therefore, the distribution has a maximum slope for a spin-singlet state~\cite{Afik:2020onf,Afik:2022kwm}.
The observable \chan is the cosine of the same angle, where the sign of the component of the lepton momentum along the top-quark flight direction is flipped~\cite{Aguilar-Saavedra:2022uye}.
Its distribution has a maximum slope for a spin-triplet state.
Both \chel and \chan exhibit different shapes in \toponium compared with \pQCD \ttbar events, as shown in Figure~\ref{fig:sel:chelchan}, which displays the parton-level distributions directly from the matrix-element calculation, prior to parton showering.
The \toponium distributions agree well with those obtained analytically for pure $^1S_0$ states. The small differences between these contributions can be attributed to a small $P$-wave contribution in the underlying \pQCD $gg\to\ttbar$ ME, which is reweighted with the NRQCD Green’s function to obtain the \toponium MC sample (Section~\ref{subsec:signal_samples}). The Green's function itself does not include $P$-wave contributions.
The impact of these differences on the measurement results was tested by reweighting the \toponium sample to the analytical predictions for pure $^1S_0$ states in the relevant angular variables. The impact was found to be negligible.

\begin{figure}[htbp]
\centering
\includegraphics[width=0.5\textwidth]{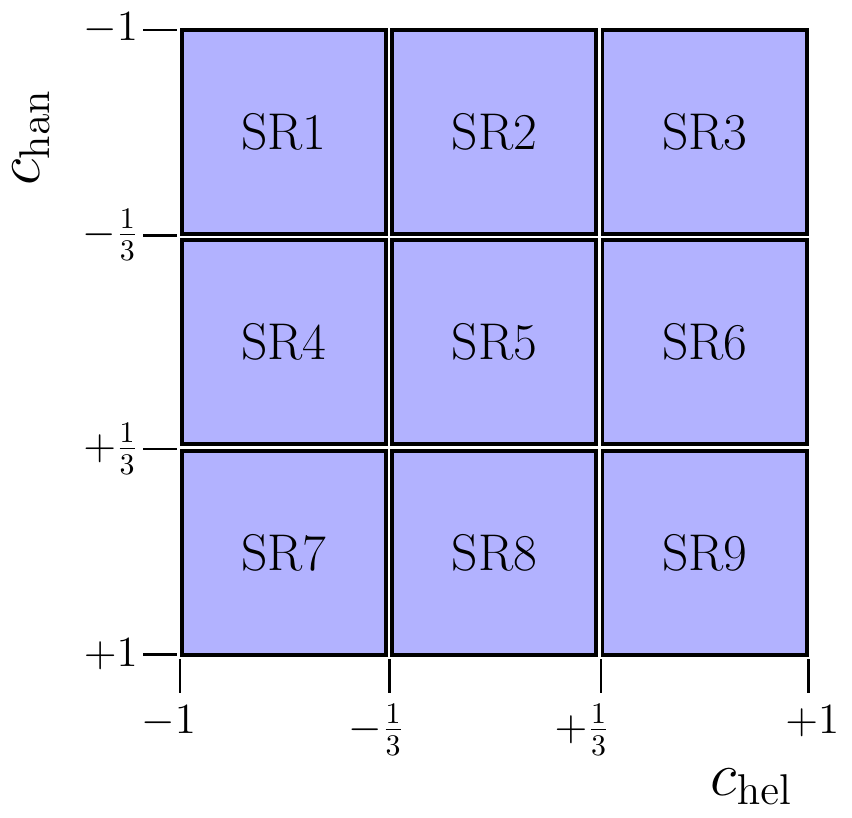}
\caption{Definition of the nine signal regions based on the angular variables \chel and \chan.}
\label{fig:results:SRsplit}
\end{figure}

For each angular observable, three bins of equal width are defined: $-1\leq \chel < -\frac{1}{3}$, $-\frac{1}{3}\leq \chel < \frac{1}{3}$, $\frac{1}{3}\leq \chel < 1$ (and similarly for \chan).
Here and in the following the SRs are numbered from 1 to 9, corresponding to the bins shown in Figure~\ref{fig:results:SRsplit}.
In each SR, the distribution of the reconstructed invariant mass of the \ttbar system, \mttbar, is analysed in four equidistant bins across the range 300 -- 500~\GeV, see Section~\ref{sec:results} for further details on the final fit setup.

%

%
\section{Reconstruction of the top-antitop system}
\label{sec:reco_2l}

The reconstruction of the \ttbar system from the two selected $b$-jet candidates, the charged leptons, and the \ptmiss is based on the Ellipse Method~\cite{Betchart:2013nba}, a geometric approach to analytically solve a system of constrained equations to obtain the four-vectors of the two neutrinos from the leptonically decaying $W$ bosons.
The two $b$-jet candidates associated with the decays of the top and antitop quarks are chosen from all selected hadronic jets. If more than two of them are $b$-tagged, the two highest-\pT $b$-tagged jets are selected. If there is only one $b$-tagged jet, the highest-\pT jet among the remaining untagged ones is selected.

The following kinematic constraints are imposed: the invariant masses of the $\ell^+\nu_{\ell}b$ and $\ell^- \bar{\nu}_{\ell}\bar{b}$ are equal to $\mtop = 172.5$~\GeV; the invariant masses of the $\ell^+\nu_{\ell}$ and $\ell^- \bar{\nu}_{\ell}$ systems are equal to the $W$-boson mass of 80.4~\GeV; and the two neutrinos $\nu_{\ell}$ and $\bar{\nu}_{\ell}$ are the only source of \ptmiss.
The constraint equations are solved for both possible lepton-$b$-jet combinations in an event.
The Ellipse Method can yield between zero and four solutions.
If there are multiple solutions, the solution with the smallest $m_{\ttbar}$ value is used.
If no solution is found for any of the lepton-$b$-jet combinations, the constraint equations are evaluated for 100 different random values of the top-quark and $W$-boson masses, with the mass values following a Gaussian distribution with mean $\mu = 172.5~(80.379)$~\GeV, and standard deviation $~\sigma = 1.48~(2.085)$~\GeV for the top quark ($W$ boson), respectively.
If several of these pseudo-experiments yield at least one solution, again the solution with the lowest value of $m_{\ttbar}$ is chosen.

The Ellipse Method provides a solution for about 95\% of \ttbar dilepton events for both the \phhvqpy and \toponium samples in the SR. Events where no solution was found are discarded. The fraction of events without a solution is consistent between data and simulation.
The resolution of the reconstructed \mttbar, $\mttbar^\text{reco}$, quantifies the migration of events between different \mttbar bins. It is taken as the standard deviation of the distribution of the quantity $(\lvert\mttbar^\text{reco}-\mttbar^\text{true}\rvert)/\mttbar^\text{true}$, where $\mttbar^\text{true}$ is the parton-level \mttbar after the emission of final-state radiation. It is evaluated in bins of the $\mttbar^\text{true}$ distribution, each with a width of 5~\GeV, and is about 22\% at the \ttbar threshold and improves to about 18\% for $\mttbar^\text{true} \approx 500$~\GeV.

%
%

%


%

%
\section{Estimation of other Standard Model processes}
%

After applying all selection requirements, the SRs are expected to contain about 650,000 \ttbar events based on the baseline \phhvqpy model before the profile-likelihood fit to data.
This number increases by around 1.1\% (+6,900 events) for the extended sample,
which includes in addition the predictions for \ttbar quasi-bound-state formation using the normalised nominal sample described in Section~\ref{subsec:ttbar_samples}.
Smaller background contributions arise from various sources.
The largest non-\ttbar background is the production of a single top-quark in association with a $W$ boson ($tW$), with a $4\%$ contribution to the total event yields.
Rare processes involving top quarks, labelled $t+X$ and including e.g. $\ttbar+Z$, $\ttbar+W$, $\ttbar+H$ and their single-top counterparts, as well as diboson ($VV+$jets) production account for $0.3\%$ and $0.2\%$ of the total pre-fit event yields, respectively. These processes are all taken from MC simulation.

Two additional background components are estimated by using MC simulation corrected with data from dedicated control regions.
The background from $Z$+jets production contributes to the SRs mostly via non-resonant Drell-Yan production and also on-shell via $Z\to\tau^+\tau^-$ where the $\tau$-leptons decay leptonically.
This contribution represents $0.8\%$ of the total number of events in the SRs.
The $Z$+jets process is separated into three components according to the flavour of the additional leading jet in the event: $Z+b$, $Z+c$ and $Z+\mathrm{light}$.
The $Z+b$ component is dominant and subject to significant modelling uncertainties in this region of phase space~\cite{STDM-2018-43}.
A normalisation factor for $Z+b$ is therefore obtained in the profile-likelihood fit from the dedicated CR-Z regions (Section~\ref{sec:selection}).
The $Z+c$ and $Z+\mathrm{light}$ components are taken directly from the MC simulations and assigned conservative normalisation uncertainties (Section~\ref{sec:systematics}).

Another small background component arises from processes with at least one fake or non-prompt lepton that  satisfies the lepton identification and isolation criteria (Section~\ref{sec:event}).
This is mostly \ttbar production with one top quark decaying hadronically and the other leptonically, as well as single-top and $W$+jets production.
It contributes around $1.5\%$ to the total event yield in the SRs.
The two main sources of fake leptons are electrons from photon conversions, with a contribution of 30-40\% to the total fake-lepton background, and electrons and muons from heavy-flavour hadron decays, accounting for most of the remaining fake-lepton events.
Electrons originating from heavy-flavour hadron decays are characterised by a softer transverse momentum spectrum compared with those produced from photon conversions.
The \mttbar shapes in the SRs for these three different fakes contributions in the SRs are taken from MC simulations.
Their normalisation is determined by fitting the \pT distributions of the trailing lepton, defined as the lower-\pT lepton, to data in the fake-lepton-enriched CR-Fakes regions (Section~\ref{sec:selection}).
Three independent normalisation factors, \efakes, \mufakes, and \phfakes, are defined, depending on whether the fake lepton is an electron or a muon from a heavy-flavour decay or an electron arising from a photon conversion, respectively.
They are free-floating in the final profile-likelihood fit.


%

%
\section{Systematic uncertainties}
\label{sec:systematics}

%
%
%

A broad range of experimental and modelling uncertainties are considered for \phhvqpy, \toponium, and non-\ttbar processes.
These can affect both the shape and the normalisation of the \mttbar spectra in the different SRs.

\subsection{Experimental uncertainties}
%

%
%
%
%
Experimental uncertainties cover reconstructed objects, pile-up modelling, and luminosity.
Most uncertainties affect both shape and normalisation, while the luminosity only affects the normalisation.
They are applied to all processes, with details provided below.

\textbf{Jets:}
The dominant jet-related uncertainties are uncertainties in the jet-energy scale (JES) and resolution (JER)~\cite{JETM-2018-05}. These include uncertainties related to the dependence of the jet response on the pseudo-rapidity range (jet $\eta$ inter-calibration~\cite{JETM-2018-05}), the jet flavour composition~\cite{ATL-PHYS-PUB-2022-021}, punch-through, calorimeter response to different jet flavours and pile-up effects. An uncertainty in the JVT efficiency is also included~\cite{PERF-2014-03}.

\textbf{$b$-tagging:}
Uncertainties in the $b$-tagging efficiency and mis-tag rates for $c$- and light-flavour jets,
derived from dedicated measurements in flavour-enriched data and MC samples~\cite{FTAG-2018-01,FTAG-2020-08,FTAG-2019-02}, are included.

\textbf{Leptons:} Uncertainties in electron and muon trigger, identification, isolation and reconstruction efficiencies as well as uncertainties related to the energy and momentum scale and resolution are also included. They are derived from dedicated analyses using data control samples~\cite{EGAM-2018-01,EGAM-2021-01,MUON-2018-03,MUON-2022-01}.

\textbf{\met:} Object-related uncertainties are propagated to \met, with additional contributions from the scale and resolution of tracks not associated with leptons or jets~\cite{JETM-2020-03}.

\textbf{Pile-up:} Uncertainties in the pile-up reweighting are included, accounting for differences between predicted and measured values of the inelastic cross-sections and of the average number of interactions per bunch crossing~\cite{PERF-2014-03}.

\textbf{Luminosity:}
A 0.83\% normalisation uncertainty in the integrated luminosity~\cite{DAPR-2021-01} is applied to all samples except the fake-lepton and $Z$+jets backgrounds, which are estimated from data.

%
%

%
%
%
%
%
%

%


\subsection{Modelling uncertainties}

\subsubsection{\pQCD \ttbar production}
\label{subsec:sys_tt}

An extensive range of uncertainties related to the MC modelling of \pQCD \ttbar production using \phhvqpy is taken into account.
They include the following uncertainties in the NNLO-QCD+NLO-EW prediction to which the generated nominal and alternative samples are reweighted.

\textbf{NNLO-QCD scales:}
The effect of missing higher-order corrections is estimated by reweighting the \phhvqpy sample to alternative NNLO-QCD+NLO-EW predictions obtained by varying $\mu_R$ and $\mu_F$ by factors of 2 and 0.5 and taking the envelope of the resulting variations. All $\mu_R$ and $\mu_F$ combinations are considered except those for which $\mu_R=2, \mu_F=0.5$ and vice versa, see Ref.~\cite{Cacciari:2011hy}.
The envelope is dominated by the $\mu_R$ variation.

\textbf{NLO-EW correction scheme:}
An additional uncertainty comes from the difference between additive and multiplicative schemes used to combine NLO EW and NNLO \pQCD corrections~\cite{Kuhn:2013zoa}.

In the Standard Model, the top-quark mass and its Yukawa coupling are directly proportional to each other. Since an uncertainty in the top-quark mass is explicitly accounted for (see below), no additional uncertainty is assigned to the Yukawa coupling.
The impact of an additional uncertainty in the NLO EW correction, associated with a variation of the top-quark Yukawa coupling by $\pm 10\%$~\cite{HIGG-2021-23}, was nevertheless evaluated and found to have a negligible effect on the results presented in Section~\ref{sec:results}.

Uncertainties related to generator and parameter choices in the nominal \phhvqpy sample are also taken into account in the fit. They are described in the following.
In the case of one-sided uncertainties, it is assumed that the effect is symmetric to obtain effective \textit{up} and \textit{down} variations compared with the nominal prediction.

\textbf{Top-quark mass:} The effect of the top-quark mass uncertainty is evaluated by comparing the \mttbar spectra obtained using the nominal sample to those generated with top quark masses of 172.0 and 173.0~\GeV.

\textbf{Recoil scheme:}
In simulating QCD radiation from $b$-quarks in top-quark decay, there is an ambiguity in the choice of the recoil particle after the first gluon emission~\cite{Brooks:2019xso}.
This affects out-of-cone radiation for $b$-tagged jets and hence the reconstructed top-quark mass. The associated uncertainty is evaluated using a \PYTHIA sample where, unlike the nominal sample, the top quark participates in the recoil~\cite{Bierlich:2022pfr}.

\textbf{Top-quark decay and off-shell effects:}
An uncertainty associated with the modelling of top-quark decays and off-shell effects arises from differences in the treatment of spin correlations in \ttbar production across generators.
This uncertainty is assessed by comparing the sum of the \ttbar sample and the $tW$ sample to an alternative sample generated with the \phbbfourlpy setup.
In this comparison, all samples use the \PYTHIA setting in which the top quark participates in the recoil after the first gluon emission.

\textbf{PDF + \alphas:}
PDF uncertainties are evaluated using the 30 eigenvectors in the PDF4LHC15~\cite{Butterworth:2015oua} prescription, symmetrising the full difference between the nominal PDF4LHC weight and the variation for each eigenvector. Alternative PDF sets corresponding to different \alphas values of 0.117 and 0.119 are also included.

\textbf{PS and hadronisation:}
The PS and hadronisation model uncertainty is estimated by comparing the nominal \POWHEG~+~\PYTHIA8 predictions to those obtained from a sample generated at NLO in \pQCD with \POWHEGBOX[v2] using the same settings as the nominal \POWHEG sample for the hard scattering, but interfaced with \HERWIG[7.2]~\cite{Bahr:2008pv,Bellm:2019zci}, using the default set of tuned parameters~\cite{Bellm:2019zci} and the \MMHT[lo] PDF set~\cite{Harland-Lang:2014zoa}. %

\textbf{ME-PS matching:}
The ME-PS matching uncertainty is estimated by comparing the nominal sample to one with the same settings but using \pthard = 1 instead of 0 in \PYTHIA~\cite{ATL-PHYS-PUB-2023-029}.

\textbf{\hdamp setting:}
The uncertainty related to the modelling of high-\pT radiation is estimated by comparing the nominal sample to one with the \hdamp parameter increased by a factor of 2~\cite{ATL-PHYS-PUB-2016-020}.

\textbf{Initial-state radiation (ISR):} Uncertainties in the ISR modelling are estimated according to the \textsc{Var~3c} set of parameters which are part of the A14 tune~\cite{ATL-PHYS-PUB-2014-021}.

\textbf{Final-state radiation (FSR):} Similarly, the uncertainty related to FSR is assessed by varying $\mu_R$ for final-state parton-shower emissions up and down by a factor of two. %

\textbf{Colour reconnection:}
The uncertainty associated with the modelling of colour reconnection is estimated using samples obtained with two different colour-reconnection models. In the "QCD-based" model (CR1) the formation of dipoles containing three quarks is enhanced, leading to increased baryon production, while in the "gluon-move" model (CR2) only the gluons are considered for the reconnection. To estimate the respective systematic effect, a third sample (CR0) with the nominal multiple parton interaction (MPI) model and a tune matching that of CR1 and CR2 is used. The colour-reconnection uncertainty is taken from the sample that yields the largest relative effect in a given bin~\cite{TOPQ-2019-01,ATL-PHYS-PUB-2017-008}.

\textbf{Underlying event:}
Uncertainties related to the modelling of the underlying event and MPI are evaluated with two samples where the settings for the $\alpha_s$ value used in the MPI and colour-reconnection range of the proton beam remnants are changed according to the \textsc{Var~1} set of parameters which are part of the A14 tune~\cite{ATL-PHYS-PUB-2014-021}.

The alternative MC samples used to define the modelling uncertainties in the nominal \phhvqpy simulation are all reweighted individually to the NNLO-QCD+NLO-EW predictions, with three exceptions.
No dedicated NNLO-QCD+NLO-EW weights are derived for the samples with alternative $m_t$ values to avoid underestimating this uncertainty by reweighting it to a higher-order prediction evaluated at a fixed top-quark mass $m_t = 172.5$~\GeV. Instead, the weights derived for the nominal sample are applied to these samples.
The same approach is used for the uncertainty in the choice of the PS+hadronisation model as it covers modelling differences that are not expected to affect the ME dynamics relevant for the reweighting.
The \phbbfourlpy sample, or more specifically its \ttbar component, used to derive the uncertainty in the modelling of the top-quark decay and off-shell effects is also not reweighted, as discussed in Section~\ref{sec:bb4l}.
In this case, the uncertainty is derived by comparing the alternative samples to the nominal \phhvqpy sample before reweighting.


\subsubsection{Quasi-bound-state effects}

%

Modelling uncertainties are taken into account for \ttbar quasi-bound-state production.
They are applied to both the main model and the simplified model discussed in Appendix~\ref{sec:app_toymodel}.

\textbf{Scale variations:}
The effect of missing higher-order corrections is estimated by varying $\mu_R$ and $\mu_F$ by factors of 2 and 0.5, and is treated as uncorrelated with those in the \pQCD \ttbar sample due to different orders at which the processes are modelled.
Varying $\mu_R$ and $\mu_F$ changes the predicted cross-section by 23\% and 2\%, respectively.
Although the overall normalisation is free-floating in the fit, these variations lead to a 0.5\% change in the acceptance, which does impact the cross-section measurement.

\textbf{PDF + \alphas:} Uncertainties based on the eigenvectors of the PDF set and the value of \alphas in the nominal PDF set are applied according to the same approach used for the \ttbar \pQCD sample as described in Section~\ref{subsec:sys_tt}, except that the more recent PDF4LHC21 prescription is used.

\textbf{PS and hadronisation, ISR, FSR:} Uncertainties related to the modelling of the PS and hadronisation as well as ISR and FSR are estimated by comparing the nominal predictions to those obtained from samples generated with alternative settings, defined in the same way as the equivalent uncertainties in the \ttbar \pQCD sample.

An additional uncertainty in the MC modelling of \toponium arises from the choice of the top-quark mass value in the Green's function.
Currently, no Green's function predictions for alternative values of the top-quark mass are available, hence this uncertainty is not considered in this measurement.
Given the fact that the overall normalisation of the \toponium contribution is free-floating and given the limited experimental resolution close to the production threshold, the main effect of this uncertainty on the measurement is expected to arise from its impact on the acceptance.

%
%
%
%

%

%


\subsubsection{Other Standard Model processes}

%

MC modelling uncertainties for sub-dominant background processes are detailed below. For $tW$ production, both shape and normalisation effects are included. For other minor backgrounds, conservative normalisation uncertainties are used, with the exception of the fake-lepton background, whose normalisation is obtained from the fit.

\textbf{Single top quark:}
The main single-top modelling uncertainty stems from interference and overlap with \ttbar, estimated by comparing the nominal $tW$ sample using diagram-removal to an alternative sample with diagram-subtraction~\cite{Frixione:2008yi,ATL-PHYS-PUB-2016-020} under identical settings.
Uncertainties from the PS and hadronisation model, ME-PS matching (\pthard), \hdamp, and top-quark mass are estimated as in the \pQCD \ttbar sample.
A 4\% normalisation uncertainty reflects the $tW$ cross-section uncertainty used to normalise the MC samples~\cite{Kidonakis:2021vob}.

\textbf{$\ttbar+b$ and $\ttbar+c$:}
Events in which a \ttbar pair is produced in association with an additional $b$- or $c$-quark, referred to as $\ttbar+b$ and $\ttbar+c$, respectively, are identified within the nominal \pQCD \ttbar sample using particle-level information.
The relative contribution of the $\ttbar+b$ ($\ttbar+c$) subsample is varied by 20\% (40\%) to account for the uncertainties in the relative cross-sections of these processes~\cite{TOPQ-2019-03,TOPQ-2021-26}.

\textbf{$\ttbar+X$:} A conservative 50\% normalisation uncertainty is applied to the $\ttbar+X$ ($X=W,Z,H$) and the tiny $t$+$X$ background components.

\textbf{Diboson:}
A conservative 50\% normalisation uncertainty is applied to the diboson background to cover possible mismodelling of extra and heavy-flavour jets~\cite{Alwall:2007fs,HIGG-2017-03}.

\textbf{$Z$+jets:}
For $Z$+jets, the $Z+b$ normalisation is a free parameter of the fit, while 50\% normalisation uncertainties are applied to the $Z+c$ and $Z+\mathrm{light}$ components.

%

%
%

%


\subsubsection{Correlation scheme}
\label{subsec:sys_corr}

Systematic uncertainties, including those arising from the limited size of the MC samples used to model the background processes, are included in the profile likelihood fit (Section~\ref{sec:results}) by constrained nuisance parameters (NPs), allowing for adjustments in the shape and normalisation of all histogram templates.

All experimental uncertainties are treated as fully correlated across samples and the analysis regions entering the fit.
The modelling uncertainties are treated as uncorrelated between the \ttbar, \toponium and non-\ttbar samples but, in the majority of cases, as fully correlated between the analysis regions.

A partial correlation scheme is applied for the NPs corresponding to the uncertainty in the PS and the uncertainty in the modelling of top-quark decays and off-shell effects.
These NPs are found to be strongly constrained to less than 50\% of their pre-fit uncertainty in the baseline-only fit to simulated data if they are treated as fully correlated between analysis regions. For these uncertainties, the original systematic variation is instead treated as partially (50\%) correlated between the different regions (SRs and CRs).
This is achieved by splitting it into one part, obtained from the original systematic variation scaled by a factor of $1/\sqrt{2}$, which is treated as fully correlated across SRs, and $N$ parts obtained in the same way but treated as fully uncorrelated across the $N=13$ regions entering the fit.
This approach was found to only slightly reduce the sensitivity of the measurement and to have a negligible effect on the normalisation factors for the backgrounds, while significantly reducing the NP constraints in the SRs and improving the Goodness-of-Fit value.
The Goodness-of-Fit value is calculated using a saturated model~\cite{Cousins:2018tiz} and quantifies the agreement between the data and the model under consideration, accounting for the penalties arising from shifts in the nuisance parameters during the fit. A high Goodness-of-Fit value indicates good agreement between the data and the model with minimal shifts in the nuisance parameters.

Alternative correlation approaches for these two uncertainties were tested, including a full correlation across all regions, a full decorrelation between the nine SRs, and a splitting of the related NPs by their shape and acceptance components.
Additionally, the partial decorrelation approach was tested for the slightly less constrained NPs, such as those related to the choice of the recoil-to-top setting.
The effects of these alternative correlation approaches on the final fit results are summarised in Section~\ref{sec:fit_results}.
%

%
%



%

%
%
%
%
%

%
\section{Results}
\label{sec:results}

The agreement between the data and the predictions of the baseline \pQCD model for \ttbar production and the extended models that include quasi-bound-state effects is quantified using a binned profile-likelihood fit~\cite{Cowan:2010js}.
In total, 13 orthogonal regions are included in this fit: the nine SRs, as well as the one CR-Z and the three CR-Fakes control regions, as defined in Section~\ref{sec:selection}.

In the SRs and CR-Z, the fitted distributions are those of the reconstructed \ttbar invariant mass in the range of $300-500~\GeV$. This range is split into four bins with equal width of $50~\GeV$.
In the CR-Fakes regions, the distribution of the trailing lepton \pT is used. In this case three bins are defined, bounded by $[10,20,40,200]~\GeV$.
The templates used in the fit are obtained from the MC simulations described in Section~\ref{sec:samples}.

Six normalisation factors  defined as $\mu = \sigma_{\textrm{meas.}}/\sigma_{\textrm{theory}}$ for the \toponium, \ttbar, \efakes, \phfakes, \mufakes, and $Z+b$ processes are extracted simultaneously from the fit.
The use of two separate normalisation factors for the \pQCD \ttbar component and the quasi-bound-state contributions makes it possible to test the level of agreement between the baseline model ($\mu(\ttbar) = 1, \mu(\toponium) = 0$) and the extended model ($\mu(\ttbar) = 1, \mu(\toponium) = 1$), smoothly interpolating between the two scenarios.
The normalisation factor $\mu(\toponium)$, also referred to as the \textit{signal strength}, is the ratio between the measured cross-section and the predicted cross-section to which the \toponium sample is normalised.

\begin{figure}[htbp]
\centering
\includegraphics[width=\textwidth]{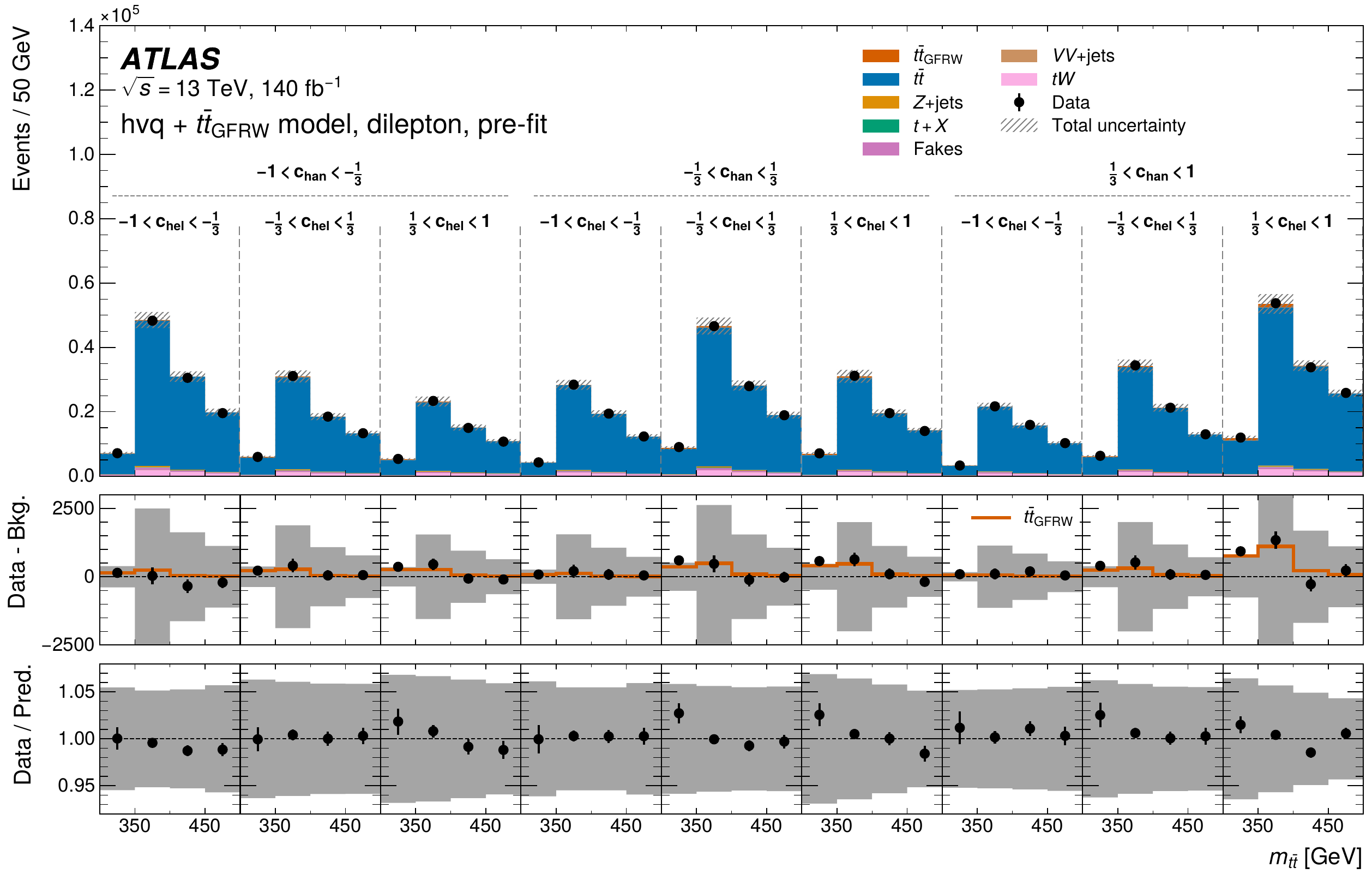}
\caption{Pre-fit distributions of \mttbar in the nine SRs (upper panel), together with a comparison between the \ttbar quasi-bound-state prediction and the data, from which the \pQCD \ttbar contribution and background processes were subtracted (middle panel), and the ratio of the data and the extended \ttbar model including the \toponium contribution (lower panel). The error bars on the data markers represent the statistical uncertainty in the measurement, while the grey hashed and shaded bands represent the total systematic uncertainty in the prediction.}
\label{fig:results:unblinded:hvq:summary2l_prefit}
\end{figure}

The \mttbar distributions in the nine SRs before the fit are shown in Figure~\ref{fig:results:unblinded:hvq:summary2l_prefit}, using the numbering scheme defined in Section~\ref{sec:selection} from left to right.
The data are above the central value of the prediction of the baseline \phhvqpy model without \toponium effects included, as visible in all SRs at low values of \mttbar, most notably in SRs~5, 6, 8, and 9, characterised by larger values of \chel and \chan.
Overall, the data are in better agreement with the predictions of the extended model that includes the quasi-bound-state \toponium predictions (normalised to 6.43~pb).
In SRs 5, 6, 8 and 9, the observed data yields in the lowest bin in \mttbar notably exceed those predicted in the \toponium model.

\subsection{Fit results}
\label{sec:fit_results}

The cross-section for the \ttbar quasi-bound-state contribution extracted from the fit with free-floating signal normalisation $\mu(\toponium)$ is
\begin{equation*}
\sigma(\toponium) = \; 9.3^{+1.4}_{-1.3}~\mathrm{pb} \; = \; 9.3^{+1.1}_{-1.0}~\mathrm{(stat.)} \pm 0.8~\mathrm{(syst.)}~\mathrm{pb} , %
\end{equation*}
which is $(45^{+21}_{-20})\%$ larger than the predicted value of 6.43 pb.
The baseline model without quasi-bound-state contributions is rejected with an observed (expected) significance of over $8\sigma$ ($6\sigma$).

The statistical uncertainty in this measured cross-section is calculated based on the requirement that the sum in quadrature of the statistical uncertainty and the total systematic uncertainty must yield the total uncertainty as obtained from the profile-likelihood fit. %
The uncertainty components from each individual source of systematic uncertainty are taken from the covariance matrix obtained from the fit and combined in quadrature to obtain the total systematic uncertainty; the statistical uncertainty is then derived as the difference in quadrature between the total and systematic uncertainties~\cite{Pinto:2023yob}. This method was found to reproduce most accurately the relative size of the statistical and systematic uncertainties, as obtained from pseudo-experiments.
An alternative method for estimating the statistical uncertainty is used in Ref.~\cite{CMS-TOP-24-007}, where the statistical component is obtained by fixing all nuisance parameters to their post-fit values, while the systematic component is computed as the quadratic difference between the total and statistical uncertainties.
Using this approach, the cross-section obtained with this measurement is $\sigma(\toponium) = 9.3 \pm 0.6~\mathrm{(stat.)}^{+1.2}_{-1.1}~\mathrm{(syst.)}~\mathrm{pb}$.

The other normalisation factors extracted from the fit with a free-floating \toponium contribution are
\begin{align*}
\mu(\ttbar) &= 1.01 \pm 0.03 \\
\mu(\efakes) &= 1.04 \pm 0.07 \\
\mu(\phfakes) &= 1.03^{+0.16}_{-0.23} \\
\mu(\mufakes) &= 1.19 \pm 0.04 \\
\mu(Z+b) &= 0.78^{+0.42}_{-0.35}
\end{align*}

As a cross-check, the normalisation factors were also fitted separately for three different data-taking periods during LHC Run~2: 2015+2016, 2017, 2018, and separately in $ee$, $e\mu$, and $\mu\mu$ events. In all cases, the results agree with those derived from the nominal fit within their uncertainties.

The stability of this result was also tested under the alternative NP correlation schemes described in Section~\ref{subsec:sys_corr}. The differences between the fitted normalisation factors compared with the nominal setup were negligible for most alternative correlation schemes and the impact on the associated uncertainties was small. The approach in which all NPs are treated as fully correlated across regions yields a 15\% reduction of the uncertainty in the normalisation factor for the \toponium process compared with the nominal correlation scheme, with a negligible change in the fitted central value.

\begin{figure}[htbp]
\centering
\includegraphics[width=\textwidth]{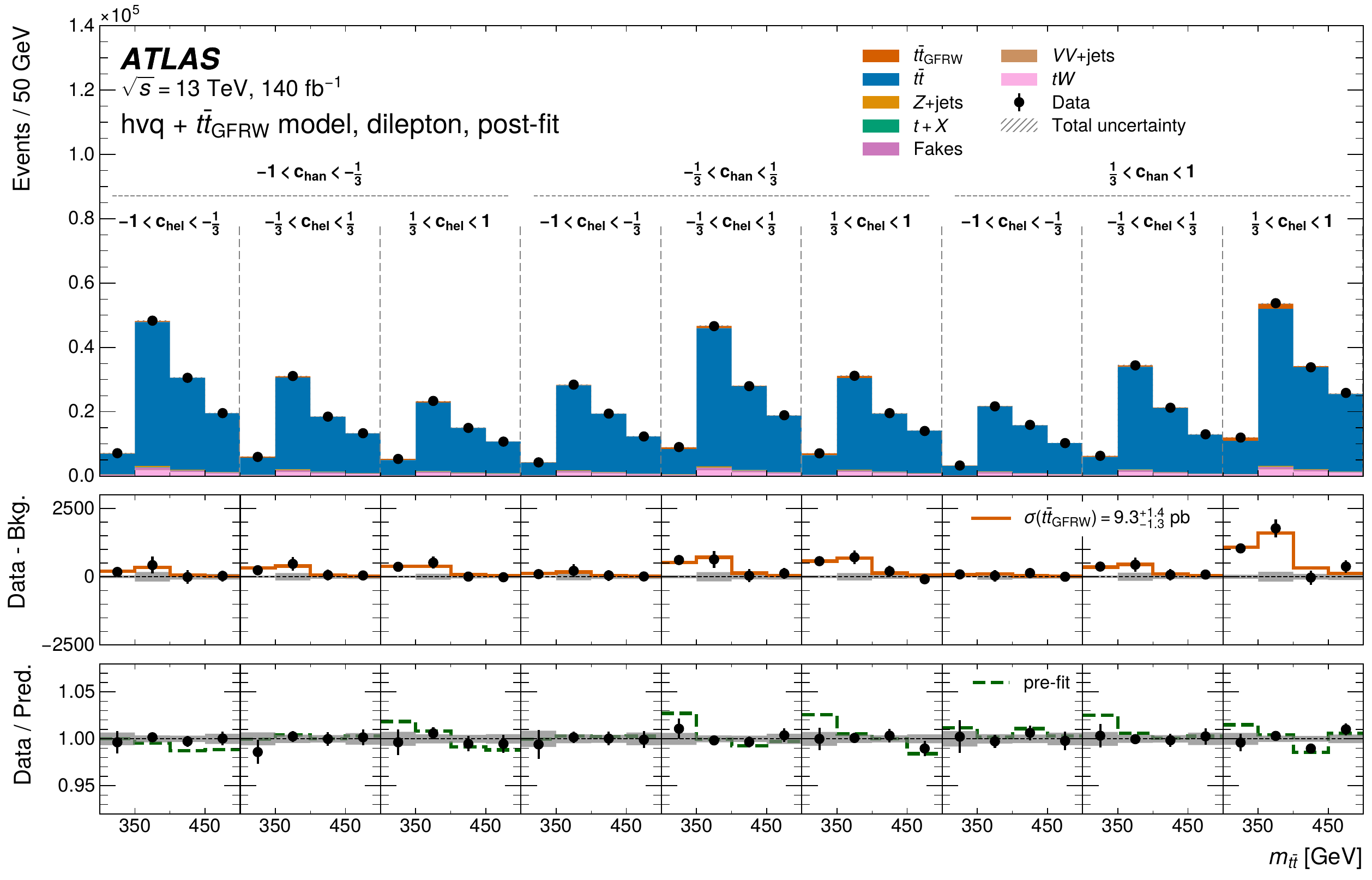}
\caption{
Post-fit distributions of \mttbar in the nine SRs (upper panel), together with a comparison between the \ttbar quasi-bound-state prediction and the data, from which the \pQCD \ttbar contribution and background processes were subtracted (middle panel), and the ratio of the data and the extended \ttbar model including the \toponium contribution (lower panel). The error bars on the data markers represent the statistical uncertainty in the measurement, while the grey hashed and shaded bands represent the total systematic uncertainty in the prediction. The dashed line represents the data/MC ratio before the fit.}
\label{fig:results:unblinded:hvq:summary2l_postfit}
\end{figure}

The \mttbar distributions in the nine SRs after this fit are shown in Figure~\ref{fig:results:unblinded:hvq:summary2l_postfit}.
Good agreement for the extended model is observed after the fit, with significantly reduced NP shifts compared with the fit without the quasi-bound-state contribution.
Accordingly, the Goodness-of-Fit value degrades from 0.98 in the fit with the extended model to smaller than $10^{-5}$ when the quasi-bound-state is omitted.
The data agree with the predictions in all CRs.

\begin{figure}[htbp]
\centering
\includegraphics[width=0.8\textwidth]{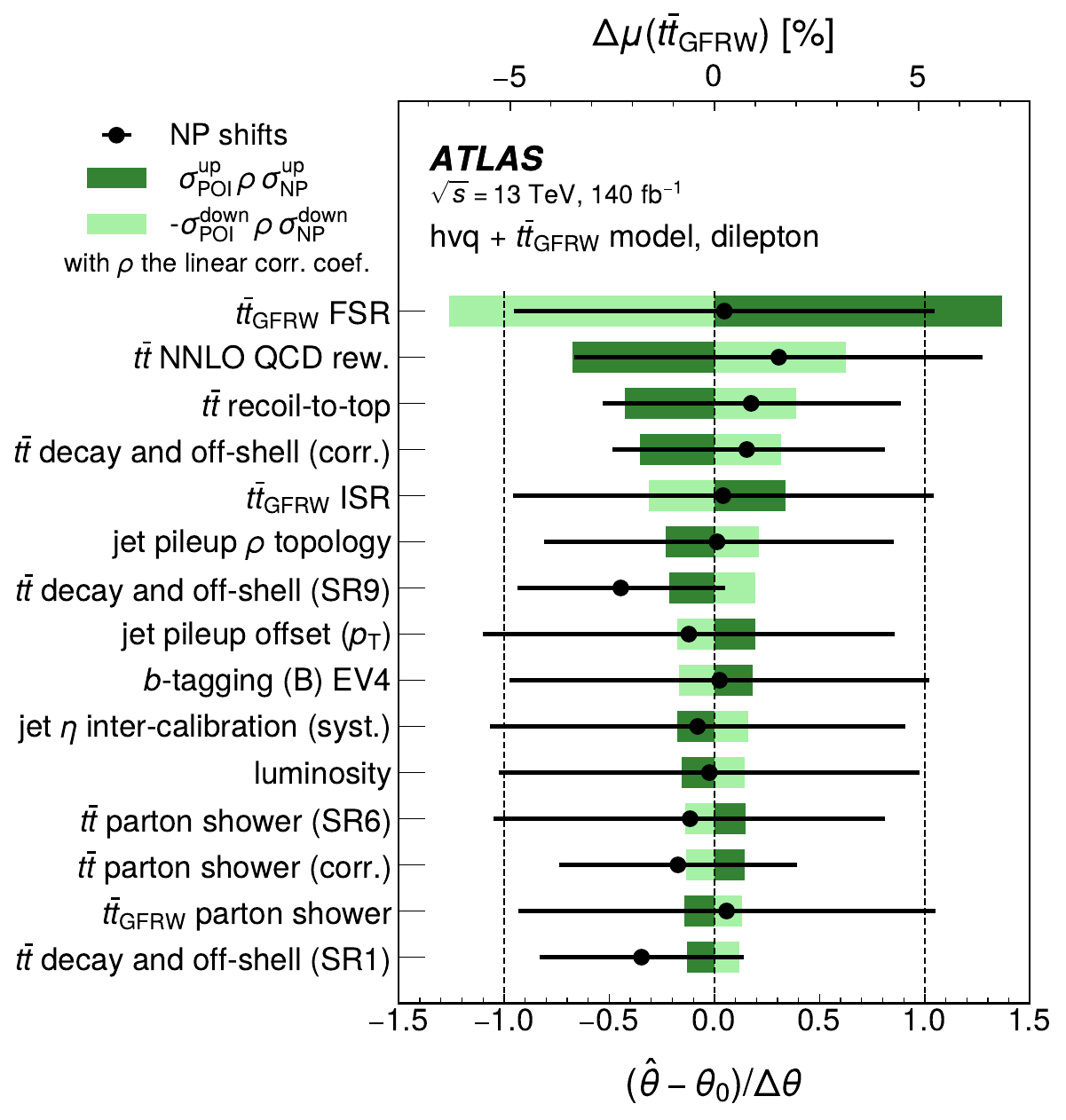}
\caption{Ranking of the 15 most impactful individual NPs on $\mu(\toponium)$ in the profile-likelihood fit with free-floating $\mu(\toponium)$.
The covariance-matrix approximation of the shifted observables method is used to evaluate the impact of a given NP~\cite{Pinto:2023yob}.
The impact is taken as the corresponding off-diagonal element of the fit covariance matrix divided by its pre-fit uncertainty and is given in percent on the upper scale. The black marker shows the shifts of the NPs relative to the nominal value $\theta_0$. They are shown together with their corresponding uncertainty on the lower scale.
The SR numbers correspond to the bins introduced in Figure~\ref{fig:results:SRsplit}. For the flavour-tagging uncertainties, individual eigenvariations (EVs) are shown.}
\label{fig:results:unblinded:hvq:ranking}
\end{figure}

Figure~\ref{fig:results:unblinded:hvq:ranking} shows the impact of individual sources of uncertainty, as well as the shift and the constraints of the associated NPs with the largest impact on the extracted \toponium cross-section.
The impact of all sources of uncertainty is summarised in Table~\ref{tab:unblinded:grouped_uncs_hvq}.
Apart from the dominant statistical uncertainties, the measurement precision is limited by \toponium and \ttbar modelling uncertainties. For \toponium, the dominant effect is the uncertainty in the modelling of FSR, which is largest in the lowest \mttbar bins and decreases towards higher \mttbar. This is also the case for the uncertainty in the modelling of ISR for \toponium. For \pQCD \ttbar production, the highest ranking NP is that related to the scale choice in the NNLO QCD reweighting.
Among the experimental uncertainties, the NPs for the main JES uncertainty components related to pile-up ("jet pile-up $\rho$ topology" and "jet pile-up offset ($\pT$)") are the highest ranking with a moderate constraint of 0.83 for the former and no constraint for the latter. The former accounts for the difference in the average pile-up energy density $\rho$ between the dijet and \ttbar topologies that affect the area-based pile-up subtraction applied to jets and hence the measured JES~\cite{JETM-2018-05}.
The latter accounts for offsets and $\pT$ dependencies in the average number of interactions per bunch crossing and the number of pile-up vertices in an event~\cite{JETM-2018-05}.
Only moderate shifts and constraints compared with their pre-fit values are observed for the highest-ranking NPs. Among these, the largest shifts of just below 0.5 pre-fit standard deviations are found for the NP on the uncorrelated parts of the uncertainty in the modelling of the \ttbar decay and off-shell effects in SR9 and SR1, respectively. This NP also exhibits the strongest constraint (0.49), while its higher-ranking, fully correlated component is less constrained (0.65).

\begin{table}[!htb]
\footnotesize
\caption{
Breakdown of the total uncertainty in the fitted normalisation factor $\mu (\toponium)$ into its statistical and systematic components.
The impact for a given NP is taken as the corresponding off-diagonal element of the fit covariance matrix divided by its pre-fit uncertainty.
The impact of a group of NPs is obtained by summing the impacts of all NPs in this category in quadrature.
The total systematic uncertainty is obtained by summing in quadrature the impacts of all NPs. The statistical uncertainty in the signal strength is calculated based
on the requirement that the sum in quadrature of the statistical uncertainty and the total systematic uncertainty must yield the total uncertainty in the signal strength as obtained from the initial profile-likelihood fit in which all NPs are allowed to float.
The uncertainties are symmetrised by taking the average of the upward and downward variations.
The category "Instrumental (other)" considers the uncertainty in the luminosity, pile-up reweighting and jet-vertex tagger efficiency.}
\label{tab:unblinded:grouped_uncs_hvq}
\def\arraystretch{1.3}
\begin{center}
\begin{tabular}{lr}
\toprule
Category & Impact \\
\midrule
\toponium modelling & 6.9\% \\
\pQCD \ttbar modelling & 4.7\% \\
JES pile-up & 1.5\% \\
$b$-tagging & 1.3\% \\
Instrumental (other) & 0.9\% \\
JES $\eta$ inter-calibration & 0.9\% \\
Jet energy scale (JES) & 0.6\% \\
Background normalisations & 0.6\% \\
Leptons & 0.6\% \\
JES flavour & 0.6\% \\
$tW$ modelling & 0.5\% \\
Limited MC statistics & 0.3\% \\
Jet energy resolution (JER) & 0.2\% \\
\midrule
Total syst. uncertainties & 8.8\% \\
Total stat. uncertainties & 11\% \\
\bottomrule
\end{tabular}
\end{center}
\end{table}

\begin{figure}[htbp]
\centering
\includegraphics[width=0.8\textwidth]{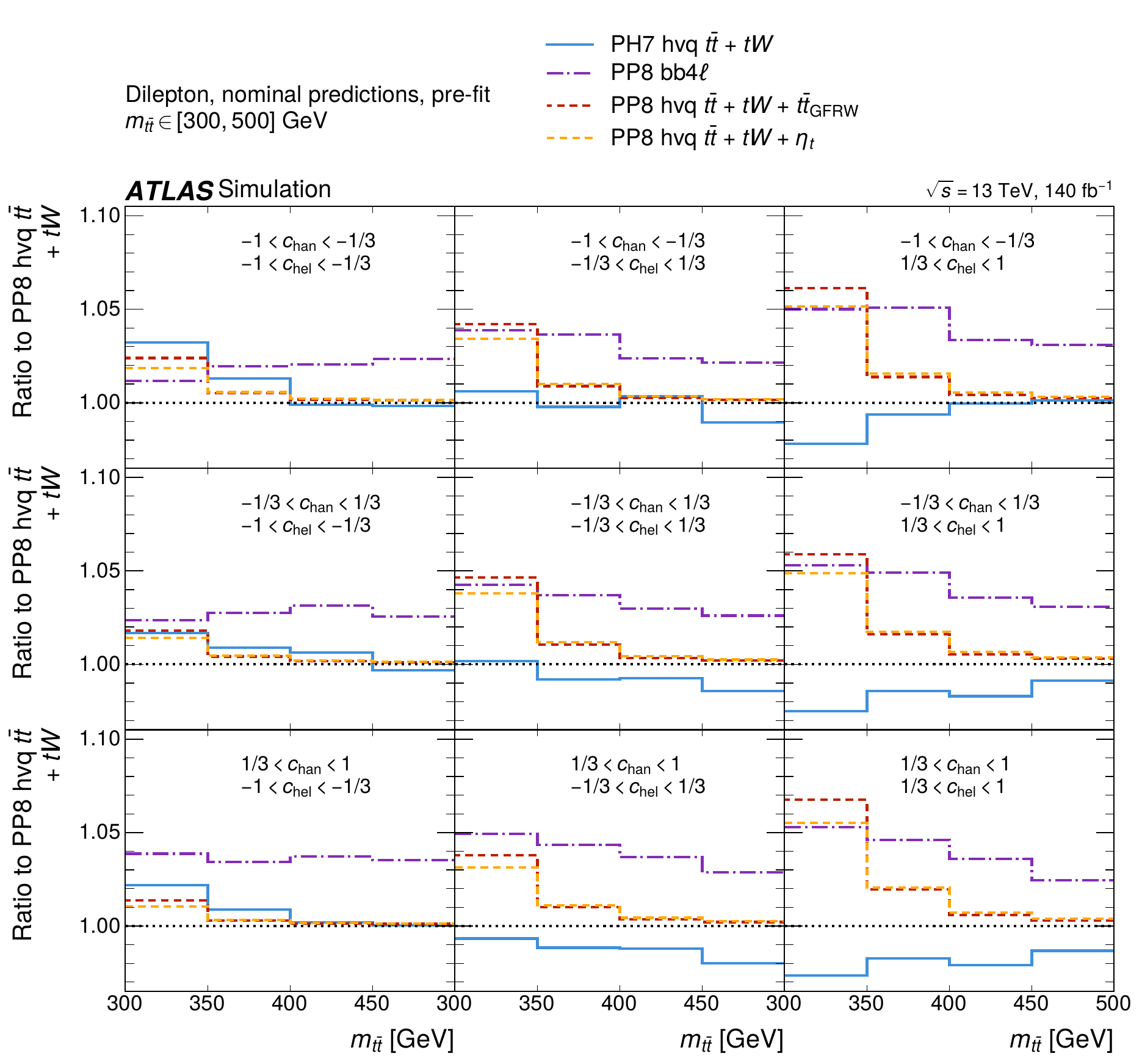}
\caption{Ratios of the pre-fit distributions of \mttbar in the nine SRs for various \ttbar models compared with the baseline \phhvqpy prediction (PP8 hvq).
The comparison is made for the extended model with the \toponium prediction and the alternative extended model using the simplified \toytoponium prediction.
In addition, the nominal baseline \phhvqpy is compared with two alternative baseline models: one using \Herwig instead of \Pythia for the PS and hadronisation (PH7 hvq) and one using \phbbfourlpy (PP8 bb$4\ell$).
The combined \ttbar+$tW$ contribution is used to ensure a consistent comparison with \phbbfourlpy.
}
\label{fig:results:summary:prefitMC:mttbar_panel}
\end{figure}

The impact of the uncertainty associated with the modelling of top-quark decays and off-shell effects can be explained via a comparison of the \mttbar distributions in the different SRs obtained with \phbbfourlpy and \phhvqpy, respectively. The $tW$ contribution was added to the \hvq-based predictions to allow for a direct comparison with the \bbfourl-based model.
In Figure~\ref{fig:results:summary:prefitMC:mttbar_panel}, the ratio between the \phbbfourlpy and \phhvqpy{}+$tW$ predictions is compared with the ratio between the \phhvqpy{}+$tW$+\toponium model and the \phhvqpy{}+$tW$ model without quasi-bound-state contributions.
In this ratio, the model with the \toponium contribution has a different shape compared with that of the \phbbfourlpy model, concentrating most of its excess compared with the \phhvqpy baseline in the two lowest \mttbar bins, while the excess predicted by \phbbfourlpy over the \phhvqpy baseline is much more evenly distributed across all four \mttbar bins. This shape difference allows the fit to distinguish a quasi-bound-state contribution from uncertainties related to the modelling of off-shell effects in the \pQCD contribution.

The predictions for the alternative extended \phhvqpy{}+\toytoponium model and the alternative baseline model using \Herwig instead of \Pythia for the PS and hadronisation are also shown in Figure~\ref{fig:results:summary:prefitMC:mttbar_panel}.
The two extended models exhibit qualitatively similar behaviour, with the \toponium model yielding a slightly larger excess of events relative to the \phhvqpy prediction in the lowest \mttbar bin than the \toytoponium model.
The \Herwig-based baseline predicts lower event yields and reduced spin correlations relative to the nominal \phhvqpy model. In contrast, both extended models predict enhanced event yields and stronger spin correlations.

An alternative set of results using \phbbfourlpy for the \pQCD \ttbar prediction is summarised in Appendix~\ref{sec:app_bb4l}.
With this setup, a cross-section of $8.5^{+1.2}_{-1.1}$~pb is measured, with a significance of over $8 \sigma$ observed ($7 \sigma$ expected).
The Goodness-of-fit value obtained for the \phbbfourlpy{}+\toponium model is 0.96, compared with a value of less than $10^{-6}$ for the fit with $\mu(\toponium)=0$.
These results are compatible with those obtained with the nominal \phhvqpy setup.
The \phbbfourlpy \pQCD setup is still under scrutiny in ATLAS, in consultation with the generator authors.
Dedicated efforts to define the best nominal settings and a corresponding systematic model are currently ongoing, including the implementation of higher-order corrections to the \ttbar and $tW$ contributions.
Further input from the theory community is needed to identify the most suitable setup.
Nevertheless, these alternative fit setups demonstrate that neither of the two baseline models of \pQCD \ttbar production describes the data well, unless a model of \ttbar quasi-bound-state formation is included. %

The results with the simplified model of Refs.~\cite{Fuks:2021xje,Maltoni:2024tul} and the \phhvqpy baseline are included in Appendix~\ref{sec:app_toymodel} to allow for a more direct comparison of this measurement with that in Ref.~\cite{CMS-TOP-24-007}.
Using the simplified model, a cross-section of $\sigma(\toytoponium) = 13.1^{+1.9}_{-1.7}$~pb is obtained, with an observed (expected) significance of over $8 \sigma$ ($4 \sigma$).
The larger observed cross-section compared with the nominal \toponium model can be explained by the differences between the \mttbar distributions in the SRs, as discussed in more detail in Appendix~\ref{sec:app_toymodel}.
Further efforts toward a more complete MC model of \nrQCD effects near the \ttbar threshold are needed to reconcile the results obtained with the simplified model (\toytoponium) and the main model (\toponium).
Such a future model should also include $P$-wave and colour-octet contributions. An improved matching between \nrQCD and \pQCD predictions is also required. Additionally, an estimate of higher-order corrections on the predicted \nrQCD effects is needed to better quantify the agreement of the measurements with the calculated \nrQCD cross-section.

%

%
%
%

%

\FloatBarrier

\section{Conclusion}
%
%

%

A measurement of \ttbar production near the kinematic threshold was conducted in events with exactly two leptons, using 140~\ifb of $\sqrt{s}=13$~\TeV $pp$ collision data collected with the ATLAS detector at the LHC.
The invariant-mass distribution of the reconstructed \ttbar system was analysed in nine orthogonal signal regions defined based on two angular variables, \chel and \chan, evaluated in the \ttbar centre-of-mass frame and sensitive to the \ttbar spin state including \toponium contributions.
The observed data were compared with two model predictions in the range of $300 < \mttbar < 500$~\GeV using a profile-likelihood fit: a baseline model including only \pQCD predictions obtained with \phhvqpy, reweighted to NNLO-QCD+NLO-EW accuracy, and an extended model that includes, in addition, state-of-the-art MC simulations of the formation of a quasi-bound-state enhancement of the cross-section near the \ttbar production threshold.
The cross-section of the additional non-relativistic QCD component, that also includes quasi-bound-state contributions, was a free parameter in the fit with the extended model.
A systematic uncertainty in the modelling of effects from off-shell top-quark decays was included via a comparison between the \phhvqpy and an alternative \phbbfourlpy prediction.

The baseline model \phhvqpy without \ttbar quasi-bound-state contributions is rejected with an observed (expected) significance of over $8 \sigma$ ($6\sigma$) due to an excess of observed events in the lowest \mttbar bins at large values of the variable \chel.
The excess over the \phhvqpy prediction is compatible with the production of colour-singlet, spin-singlet $S$-wave quasi-bound $t\bar t$ states, as expected from non-relativistic QCD, and corresponds to an observed cross-section of $9.3^{+1.4}_{-1.3}$ pb.

Further investigations are required to characterise this excess and to better quantify the impact of off-shell top-quark decays and the resummation of higher-order terms in $\alpha_s/v$ close to the threshold that are not included in nominal \phhvqpy \pQCD prediction but are expected to enhance the cross-section for \ttbar production close to threshold~\cite{Nason:2025hix}, compared with the baseline model used here. In this context, extended MC models incorporating additional \nrQCD effects, beyond $S$-wave contributions, such as higher $J$-states and octet contributions, as well as higher-order contributions and an improved matching between \nrQCD and \pQCD contributions will also be helpful to investigate the nature of this excess. %
The improved modelling is also important to search for physics beyond the Standard Model in this kinematic region~\cite{Fuks:2025toq,Flacke:2025dwk}.


%
%
%
%
%
\section*{Acknowledgements}
%

%

%
%

%
%

We thank CERN for the very successful operation of the LHC and its injectors, as well as the support staff at
CERN and at our institutions worldwide without whom ATLAS could not be operated efficiently.

The crucial computing support from all WLCG partners is acknowledged gratefully, in particular from CERN, the ATLAS Tier-1 facilities at TRIUMF/SFU (Canada), NDGF (Denmark, Norway, Sweden), CC-IN2P3 (France), KIT/GridKA (Germany), INFN-CNAF (Italy), NL-T1 (Netherlands), PIC (Spain), RAL (UK) and BNL (USA), the Tier-2 facilities worldwide and large non-WLCG resource providers. Major contributors of computing resources are listed in Ref.~\cite{ATL-SOFT-PUB-2026-001}.

We gratefully acknowledge the support of ANPCyT, Argentina; YerPhI, Armenia; ARC, Australia; BMWFW and FWF, Austria; ANAS, Azerbaijan; CNPq and FAPESP, Brazil; NSERC, NRC and CFI, Canada; CERN; ANID, Chile; CAS, MOST and NSFC, China; Minciencias, Colombia; MEYS CR, Czech Republic; DNRF and DNSRC, Denmark; IN2P3-CNRS and CEA-DRF/IRFU, France; SRNSFG, Georgia; BMFTR, HGF and MPG, Germany; GSRI, Greece; RGC and Hong Kong SAR, China; ICHEP and Academy of Sciences and Humanities, Israel; INFN, Italy; MEXT and JSPS, Japan; CNRST, Morocco; NWO, Netherlands; RCN, Norway; MNiSW, Poland; FCT, Portugal; MNE/IFA, Romania; MSTDI, Serbia; MSSR, Slovakia; ARIS and MVZI, Slovenia; DSI/NRF, South Africa; MICIU/AEI, Spain; SRC and Wallenberg Foundation, Sweden; SERI, SNSF and Cantons of Bern and Geneva, Switzerland; NSTC, Taipei; TENMAK, T\"urkiye; STFC/UKRI, United Kingdom; DOE and NSF, United States of America.

Individual groups and members have received support from BCKDF, CANARIE, CRC and DRAC, Canada; CERN-CZ, FORTE and PRIMUS, Czech Republic; COST, ERC, ERDF, Horizon 2020 and Marie Sk{\l}odowska-Curie Actions, European Union; Investissements d'Avenir Labex, Investissements d'Avenir Idex and ANR, France; DFG and AvH Foundation, Germany; Herakleitos, Thales and Aristeia programmes co-financed by EU-ESF and the Greek NSRF, Greece; BSF-NSF and MINERVA, Israel; NCN and NAWA, Poland; La Caixa Banking Foundation, CERCA and AGAUR programs from Generalitat de Catalunya and PROMETEO and GenT Programmes Generalitat Valenciana, Spain; G\"{o}ran Gustafssons Stiftelse, Sweden; The Royal Society and Leverhulme Trust, United Kingdom; Eric and Wendy Schmidt Fund for Strategic Innovation, United States of America.

In addition, individual members wish to acknowledge support from Chile: Agencia Nacional de Investigaci\'on y Desarrollo (ANID FONDECYT reg. 1230987, FONDECYT 1230812, FONDECYT 1240864, Fondecyt 3240661, Fondecyt Regular 1240721); China: Chinese Ministry of Science and Technology (MOST-2023YFA1605700, MOST-2023YFA1609300), National Natural Science Foundation of China (NSFC - 12175119, NSFC 12275265); Czech Republic: Czech Science Foundation (GACR - 24-11373S), Ministry of Education Youth and Sports (ERC-CZ-LL2327, FORTE CZ.02.01.01/00/22\_008/0004632), PRIMUS Research Programme (PRIMUS/21/SCI/017); EU: H2020 European Research Council (ERC - 101002463); European Union: European Research Council (BARD No. 101116429, ERC - 948254, ERC 101089007), European Regional Development Fund (HE COFUND GA No.101081355, ERDF), European Union, Future Artificial Intelligence Research (FAIR-NextGenerationEU PE00000013), Marie Sklodowska-Curie Actions (GAP-101168829); France: Agence Nationale de la Recherche (ANR-21-CE31-0013, ANR-21-CE31-0022, ANR-22-EDIR-0002, ANR-24-CE31-0504-01); Germany: Deutsche Forschungsgemeinschaft (DFG - 469666862); China: Research Grants Council (GRF); Italy: Ministero dell'Università e della Ricerca (NextGenEU 153D23001490006 M4C2.1.1, NextGenEU I53D23000820006 M4C2.1.1, NextGenEU I53D23001490006 M4C2.1.1, SOE2024\_0000023); Japan: Japan Society for the Promotion of Science (JSPS KAKENHI  JP25H0063, JSPS KAKENHI JP22H01227, JSPS KAKENHI JP22H04944, JSPS KAKENHI JP22KK0227, JSPS KAKENHI JP24K23939, JSPS KAKENHI JP24KK0251, JSPS KAKENHI JP25H00650, JSPS KAKENHI JP25H01291, JSPS KAKENHI JP25K01023); Norway: Research Council of Norway (RCN-314472); Poland: Ministry of Science and Higher Education (IDUB AGH, POB8, D4 no 9722), Polish National Science Centre (NCN 2021/42/E/ST2/00350, NCN OPUS 2023/51/B/ST2/02507, NCN OPUS nr 2022/47/B/ST2/03059, NCN UMO-2019/34/E/ST2/00393, UMO-2022/47/O/ST2/00148, UMO-2023/49/B/ST2/04085, UMO-2023/51/B/ST2/00920, UMO-2024/53/N/ST2/00869); Spain: Agència de Gestió d'Ajuts Universitaris i de Recerca. (AGAUR - 2023 BP 00141), Generalitat Valenciana (ASFAE/2022/008), Ministry of Science and Innovation (RYC2019-028510-I, RYC2020-030254-I, RYC2021-031273-I, RYC2022-038164-I), Ministerio de Ciencia, Innovación y Universidades/Agencia Estatal de Investigaci\'on (PID2022-142604OB-C22); Sweden: Carl Trygger Foundation (Carl Trygger Foundation CTS 22:2312), Swedish Research Council (Swedish Research Council 2023-04654, VR 2021-03651, VR 2022-03845, VR 2022-04683, VR 2023-03403, VR 2024-05451), Knut and Alice Wallenberg Foundation (KAW 2018.0458, KAW 2023.0366); Switzerland: Swiss National Science Foundation (SNSF - PCEFP2\_194658); United Kingdom: The Binks Trust, Royal Society (NIF-R1-231091); United States of America: U.S. Department of Energy (ECA DE-AC02-76SF00515), John Templeton Foundation (John Templeton Foundation 63206), Neubauer Family Foundation.

%
%


%
\clearpage
\appendix
\part*{Appendix}
\addcontentsline{toc}{part}{Appendix}
\section{Alternative fit with \phbbfourlpy}
\label{sec:app_bb4l}
%

An alternative model for \pQCD \ttbar production based on \phbbfourlpy instead of \phhvqpy was also considered in this analysis. The results obtained with this alternative model for \pQCD \ttbar production, along with the nominal \toponium model for \ttbar quasi-bound-state formation are summarised in the following.

The uncertainties in the \phbbfourlpy setup are the same as for the \phhvqpy model, with two exceptions.
First, the uncertainty in the modelling of the top-quark decay and off-shell effects, which in the case of the \phhvqpy model was derived from a comparison between \phhvqpy and \phbbfourlpy, is not applicable here.
Second, the uncertainty related to the scale choice in the NNLO QCD reweighting and the NLO-EW correction scheme, considered for \phhvqpy, is not applicable either, given no higher-order corrections to the kinematic distributions are applied to the \phbbfourlpy nominal sample, as explained in Section~\ref{sec:bb4l}.
Instead, an uncertainty due to missing higher-order corrections in the NLO QCD \phbbfourlpy prediction is derived by varying separately the renormalisation and factorisation scales of the \phbbfourlpy setup.

An uncertainty due to missing higher-order corrections in the \ttbar component of \phbbfourlpy could, in principle, also be estimated by comparing the predictions of reweighted and unreweighted \phhvqpy samples and propagating the difference to the \ttbar component of the \phbbfourlpy nominal sample. While this approach does not avoid the inaccuracies related to identifying the \ttbar component in \phbbfourlpy samples, it provides an approximation of the uncertainty that arises from not applying higher-order corrections to the kinematic distributions in the \phbbfourlpy sample. This uncertainty was found to be smaller than the uncertainty due to the variations in the renormalisation and factorisation scales of the \phbbfourlpy setup. It is therefore assumed that the uncertainty due to missing higher-order corrections is covered by these scale variations and no additional uncertainty related to missing higher-order corrections is considered for the \phbbfourlpy setup. It was verified that the results presented in this appendix change only minimally if such an additional uncertainty is included.

The other uncertainties are derived by taking the relative variations in the \phhvqpy model and applying them to the \phbbfourlpy nominal prediction.
No higher-order corrections to the kinematic distributions (Section~\ref{sec:hvq}) are applied to \phhvqpy samples in this uncertainty estimate, given that none are applied to the \phbbfourlpy nominal sample.
The same correlation scheme as for the \phhvqpy is used (Section~\ref{sec:results}).

The \mttbar distributions in the nine SRs before the fit are shown in
Figure~\ref{fig:results:unblinded:bb4l:summary2l_prefit}.
A deficit in the predicted event yields compared with the data is observed across all SRs and \mttbar bins, even if the contribution of \toponium is included. The deficit is more pronounced in the higher \mttbar bin. Overall, the pre-fit agreement between data and prediction is poorer than in the case of the nominal \phhvqpy setup (Section~\ref{sec:results}), though the distributions are still compatible within the total pre-fit uncertainty.

\begin{figure}[htbp]
\centering
\includegraphics[width=\textwidth]{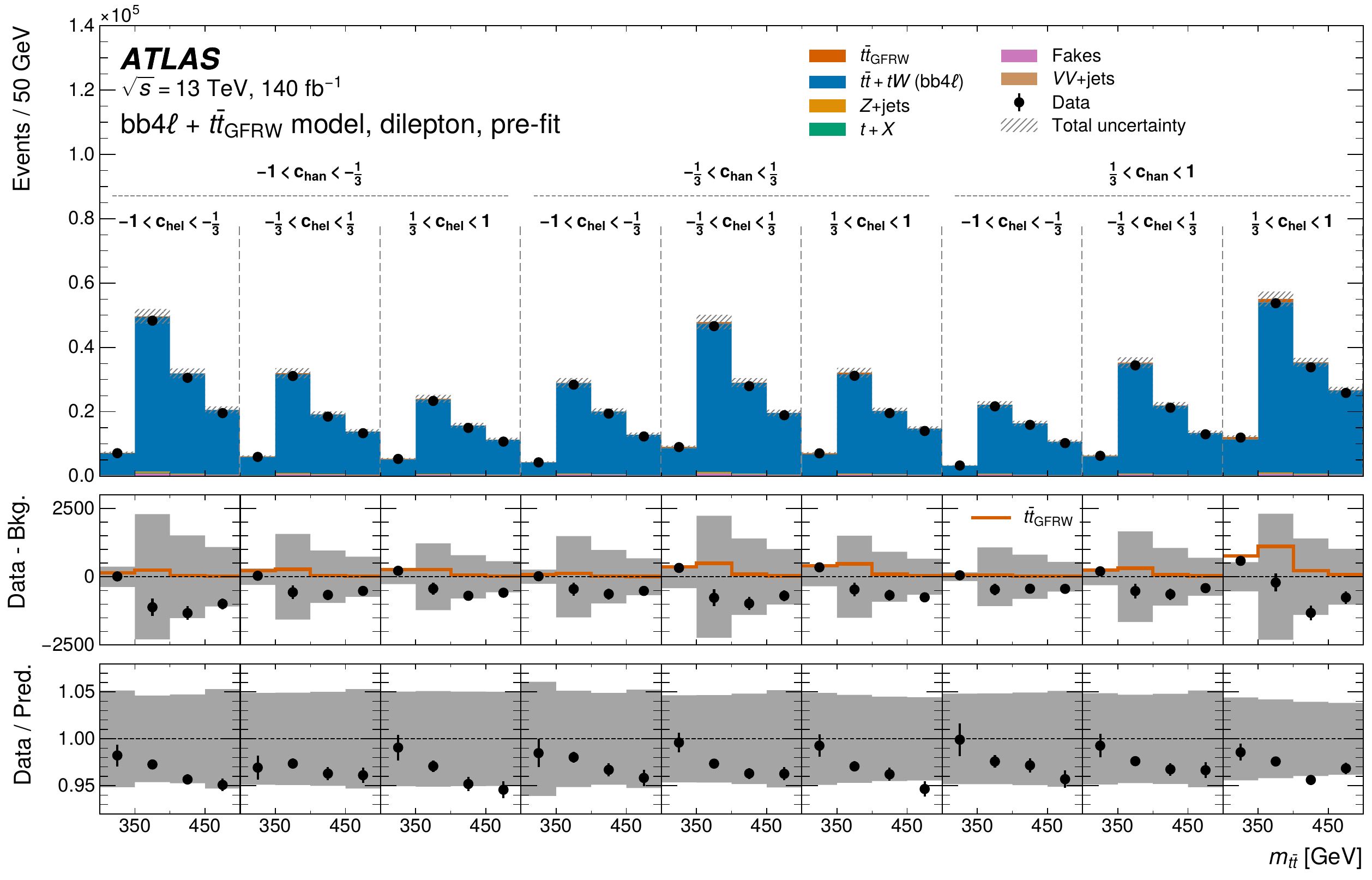}
\caption{Pre-fit distributions of \mttbar in the nine SRs (upper panel), together with a comparison between the \ttbar quasi-bound-state prediction and the data, from which the \pQCD \ttbar contribution and background processes were subtracted (middle panel), and the ratio of the data and the extended \ttbar model including the \toponium contribution (lower panel) for the \phbbfourlpy model. The error bars on the data markers represent the statistical uncertainty in the measurement, while the grey hashed and shaded bands represent the total systematic uncertainty in the prediction.}
\label{fig:results:unblinded:bb4l:summary2l_prefit}
\end{figure}

The corresponding distributions after the profile-likelihood fit to data are shown in Figure~\ref{fig:results:unblinded:bb4l:summary2l_postfit}.
Good agreement is observed between the data and the predictions of the extended model using \phbbfourlpy for the \pQCD \ttbar predictions after the fit.
The Goodness-of-Fit is 0.96, compared with a value smaller than $10^{-6}$ for the fit with $\mu(\toponium)=0$.
These values are similar to the ones for the equivalent fit with the \phhvqpy baseline model.

\begin{figure}[htbp]
\centering
\includegraphics[width=\textwidth]{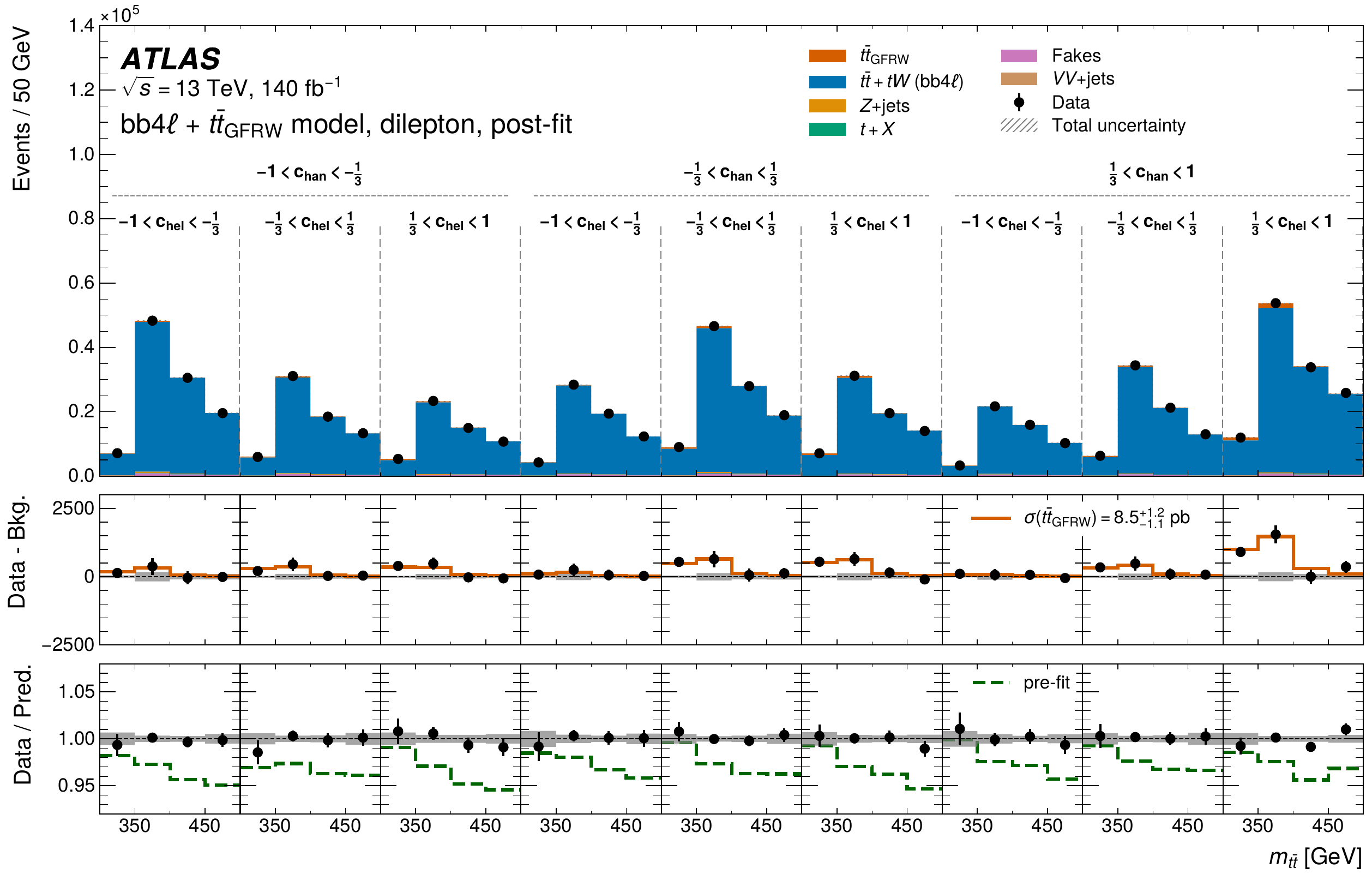}
\caption{Post-fit distributions of \mttbar in the nine SRs (upper panel), together with a comparison between the \ttbar quasi-bound-state prediction and the data, from which the \pQCD \ttbar contribution and background processes were subtracted (middle panel), and the ratio of the data and the extended \ttbar model including the \toponium contribution (lower panel) for the \phbbfourlpy model. The error bars on the data markers represent the statistical uncertainty in the measurement, while the grey hashed and shaded bands represent the total systematic uncertainty in the prediction.}
\label{fig:results:unblinded:bb4l:summary2l_postfit}
\end{figure}

The cross-section for the \ttbar quasi-bound-state sample extracted in the fit to data is
\begin{equation*}
\sigma(\toponium) = 8.5^{+1.2}_{-1.1} =  8.5^{+0.9}_{-0.8}~\mathrm{(stat.)} \pm 0.7~\mathrm{(syst.)}~\mathrm{pb}.  %
\end{equation*}
The null hypothesis $\mu(\toponium)=0$ is rejected with an observed (expected) significance of over $8 \sigma$ ($7 \sigma$). These values are comparable with those obtained for the \phhvqpy baseline.
Using the same method as in Ref.~\cite{CMS-TOP-24-007} to derive the statistical component, the resulting cross-section is $\sigma(\toponium) = 8.5 \pm 0.6~\mathrm{(stat.)}^{+1.0}_{-0.9}~\mathrm{(syst.)}~\mathrm{pb}$.

The other normalisation factors extracted from the fit with a free-floating \toponium contribution are
\begin{align*}
\mu(\ttbar+tW) &= 0.96 \pm 0.03 \\
\mu(\efakes) &= 1.05^{+0.08}_{-0.07} \\
\mu(\phfakes) &= 1.00^{+0.18}_{-0.25} \\
\mu(\mufakes) &= 1.19 \pm 0.04 \\
\mu(Z+b) &= 0.80^{+0.40}_{-0.34}
\end{align*}
The normalisation factor $\mu(\ttbar+tW)$ is about 4\% smaller than unity. This is consistent with the cross-section for off-shell top-quark pair production and decays reported in Eq.~(92) of Ref.~\cite{Buonocore:2025fqs}, which is obtained at NNLO QCD accuracy with the dipole approximation and is about 4.5\% below the sum of the higher-order \ttbar and $tW$ cross-sections used to normalise the \phbbfourlpy sample (Section~\ref{sec:bb4l}).

The systematic uncertainties with the largest impact on the extracted \toponium cross-section are shown in Figure~\ref{fig:results:unblinded:bb4l:ranking}, and the impact of all uncertainties is summarised in Table~\ref{tab:bb4l:unblinded:grouped_uncs_bb4l}.
Like in the fit based on \phhvqpy (Section~\ref{sec:results}), among the NPs with the largest impact on the extracted \toponium cross-section are those related to the modelling of FSR and ISR in the \toponium sample, and the NP for the recoil-to-top uncertainty in the \pQCD \ttbar sample.
Additionally, the NPs related to the renormalisation and factorisation scale variations in the \phbbfourlpy sample, respectively, are also high-ranking, replacing the NP related to the scale choice in the NNLO QCD reweighting for \phhvqpy, which is not applicable here.
These two NPs are also among the NPs with the largest downward shifts compared with their pre-fit values, amounting to slightly less than one standard deviation and half a standard deviation, respectively. The largest upward shift, close to one standard deviation, is observed in the high-ranking NP related to the recoil-to-top uncertainty in the \pQCD \ttbar sample.
Most NPs show only moderate constraints compared with their pre-fit values.
The strongest constraint (0.45) is observed for the correlated component of the PS uncertainty. All other NPs show constraints weaker than 0.5 of their pre-fit standard deviation.

\begin{figure}[htbp]
\centering
\includegraphics[width=0.8\textwidth]{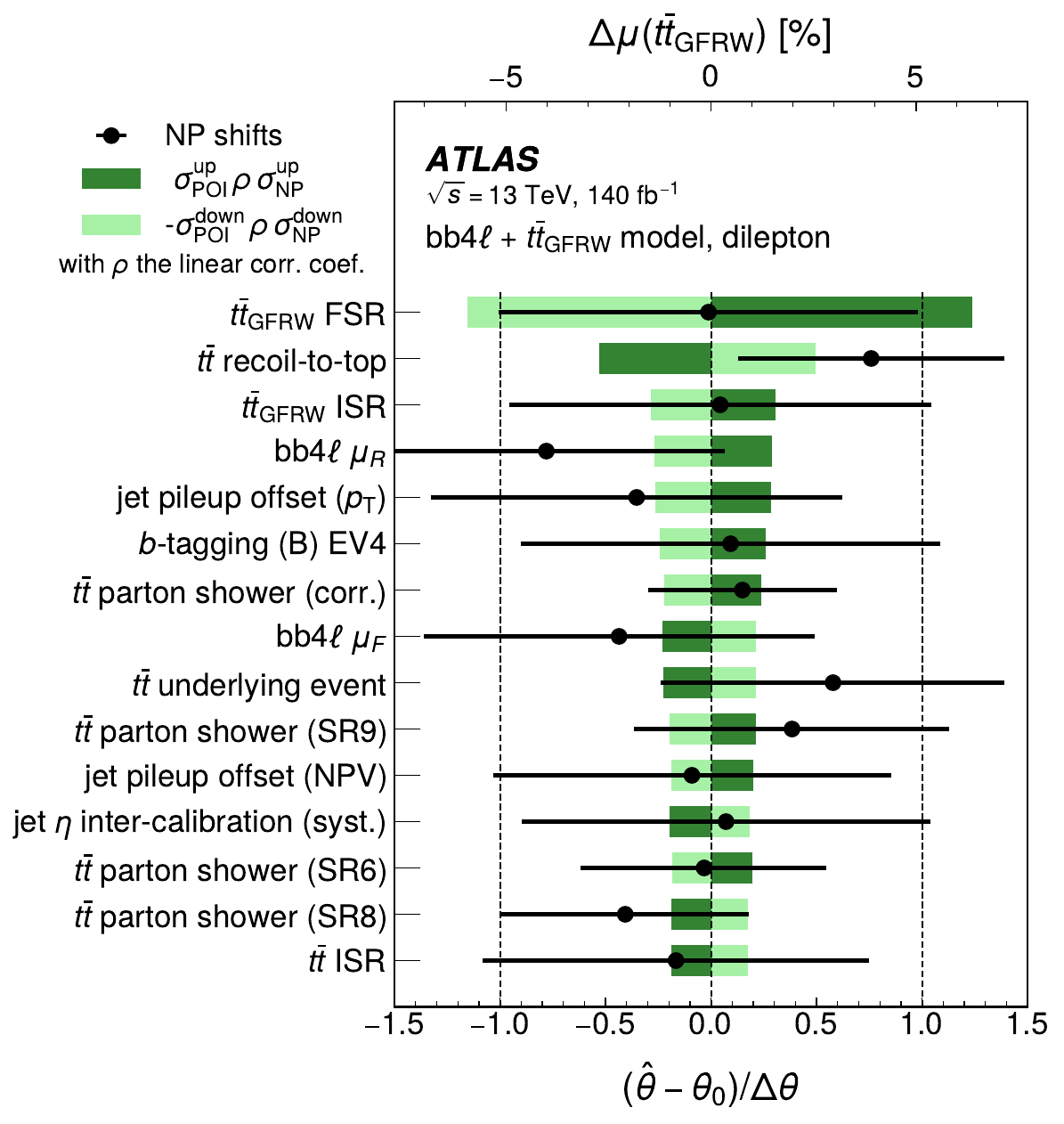}
\caption{Ranking of the 15 most impactful individual NPs on $\mu(\toponium)$ in the profile-likelihood fit with free-floating $\mu(\toponium)$.
The impact for a given NP is taken as the corresponding off-diagonal element of the fit covariance matrix divided by its pre-fit uncertainty~\cite{Pinto:2023yob} and given in percent on the upper scale. The black marker shows the shifts of the NPs relative to the nominal value $\theta_0$. They are shown together with their corresponding uncertainty on the lower scale.
The SR numbers correspond to the bins introduced in Figure~\ref{fig:results:SRsplit}.
For the flavour-tagging uncertainties, individual eigenvariations (EVs) are shown.
}
\label{fig:results:unblinded:bb4l:ranking}
\end{figure}

\begin{table}[!htb]
\footnotesize
\caption{
Breakdown of the total uncertainty in the fitted normalisation factor $\mu (\toponium)$ into its statistical and systematic components for the model based on \phbbfourlpy.
The impact for a given NP is taken as the corresponding off-diagonal element of the fit covariance matrix divided by its pre-fit uncertainty.
The impact of a group of NPs is obtained by summing the impacts of all NPs in this category in quadrature.
The total systematic uncertainty is obtained by summing in quadrature the impacts of all NPs. The statistical uncertainty in the signal strength is calculated based
on the requirement that the sum in quadrature of the statistical uncertainty and the total systematic uncertainty must yield the total uncertainty in the signal strength as obtained from the initial profile likelihood fit in which all nuisance
parameters are allowed to float.
The uncertainties are symmetrised by taking the average of the upward and downward variations.
The category "Instrumental (other)" considers the uncertainty in the luminosity, pile-up reweighting and jet-vertex tagger efficiency.}
\label{tab:bb4l:unblinded:grouped_uncs_bb4l}
\def\arraystretch{1.3}
\begin{center}
\begin{tabular}{lr}
\toprule
Category & Impact \\
\midrule
\toponium modelling & 6.3\% \\
\pQCD \ttbar modelling & 4.3\% \\
JES pile-up & 1.8\% \\
$b$-tagging & 1.6\% \\
Instrumental (other) & 1.0\% \\
JES $\eta$ inter-calibration & 1.0\% \\
Leptons & 1.0\% \\
Jet energy resolution (JER) & 0.8\% \\
JES flavour & 0.8\% \\
Background normalisations & 0.7\% \\
Limited MC statistics & 0.4\% \\
Jet energy scale (JES) & 0.2\% \\
\midrule
Total syst. uncertainties & 8.4\% \\
Total stat. uncertainties & 10\% \\
\bottomrule
\end{tabular}
\end{center}
\end{table}
\FloatBarrier


%
%
%
%
%
%
%
\clearpage
\section{Alternative fit with a simplified model of quasi-bound-state effects}
\label{sec:app_toymodel}
%

The profile-likelihood fit to data used to extract the cross-section for \ttbar quasi-bound-state production is performed also with the simplified model described in Section~\ref{subsec:signal_samples}. This provides a more direct comparison with the results of Ref.~\cite{CMS-TOP-24-007}, which reports a cross-section $\sigma(\toytoponium) = 8.8^{+1.2}_{-1.4}$~pb with this model.

The cross-section for \ttbar quasi-bound-state production extracted from the fit with the simplified model is
\begin{equation*}
\sigma(\toytoponium) = 13.1^{+1.9}_{-1.7}~\mathrm{pb} = 13.1^{+1.5}_{-1.3}~\mathrm{(stat.)} \pm 1.2~\mathrm{(syst.)~\mathrm{pb}} .
\end{equation*}
Using the same method as in Ref.~\cite{CMS-TOP-24-007} to derive the statistical component, the resulting cross-section is $\sigma(\toytoponium) = 13.1 \pm 0.8~\mathrm{(stat.)}^{+1.7}_{-1.6}~\mathrm{(syst.)}~\mathrm{pb}$.
This is roughly a factor of two larger than the predicted value of 6.43~pb.
The null hypothesis $\mu(\toytoponium)=0$ is rejected with an observed (expected) significance of over $8 \sigma$ ($4 \sigma$).
The smaller expected rejection significance compared with the nominal \toponium model is understood to originate from small differences in the \mttbar shape between the two models. As shown in Figure~\ref{fig:results:summary:prefitMC:mttbar_panel}, the simplified \toytoponium model predicts a smaller excess compared with the \phhvqpy baseline than the nominal \toponium model in the lowest \mttbar bin but a comparable relative deviation in the second \mttbar bin in each of the SRs. This leads to a smaller expected rejection significance as the relative difference between the extended model with \toytoponium compared with the baseline \phhvqpy model is smaller than for the nominal \toponium model.
The larger observed cross-section compared with the nominal \toponium model (Section~\ref{sec:results}) can also be explained by the differences between the \mttbar distributions in the SRs. To accommodate the excess in the data, which is largest in the lowest \mttbar bins, the post-fit value of the normalisation factor for the simplified model is larger than for the nominal model. The correspondingly larger post-fit yield for the simplified model in the second \mttbar bins is counterbalanced by the fit shifting the much larger \phhvqpy background, which exhibits the largest yields in the second \mttbar bins, slightly downward within its systematic uncertainties.

The other normalisation factors extracted from the fit with free-floating \toytoponium contribution are
\begin{align*}
\mu(\ttbar) &= 1.01^{+0.04}_{-0.03} \\
\mu(\efakes) &= 1.04 \pm 0.07 \\
\mu(\phfakes) &= 1.03^{+0.16}_{-0.23} \\
\mu(\mufakes) &= 1.19 \pm 0.04 \\
\mu(Z+b) &= 0.79^{+0.42}_{-0.35}
\end{align*}

The \mttbar distributions in the nine SRs before and after this fit are shown in Figures~\ref{fig:toy:unblinded:hvq:summary2l_prefit} and~\ref{fig:toy:unblinded:hvq:summary2l_postfit}, respectively.
Good agreement is observed between the data and the predictions based on \phhvqpy and the simplified model for quasi-bound-state production, with a Goodness-of-Fit value of 0.99. Similar to the main result presented in the paper body, the contribution of quasi-bound-state effects is largest in the bin with $\frac{1}{3} < \chel < 1$, $\frac{1}{3} < \chan < 1$ and at low values of \mttbar.
\begin{figure}[htbp]
\centering
\includegraphics[width=\textwidth]{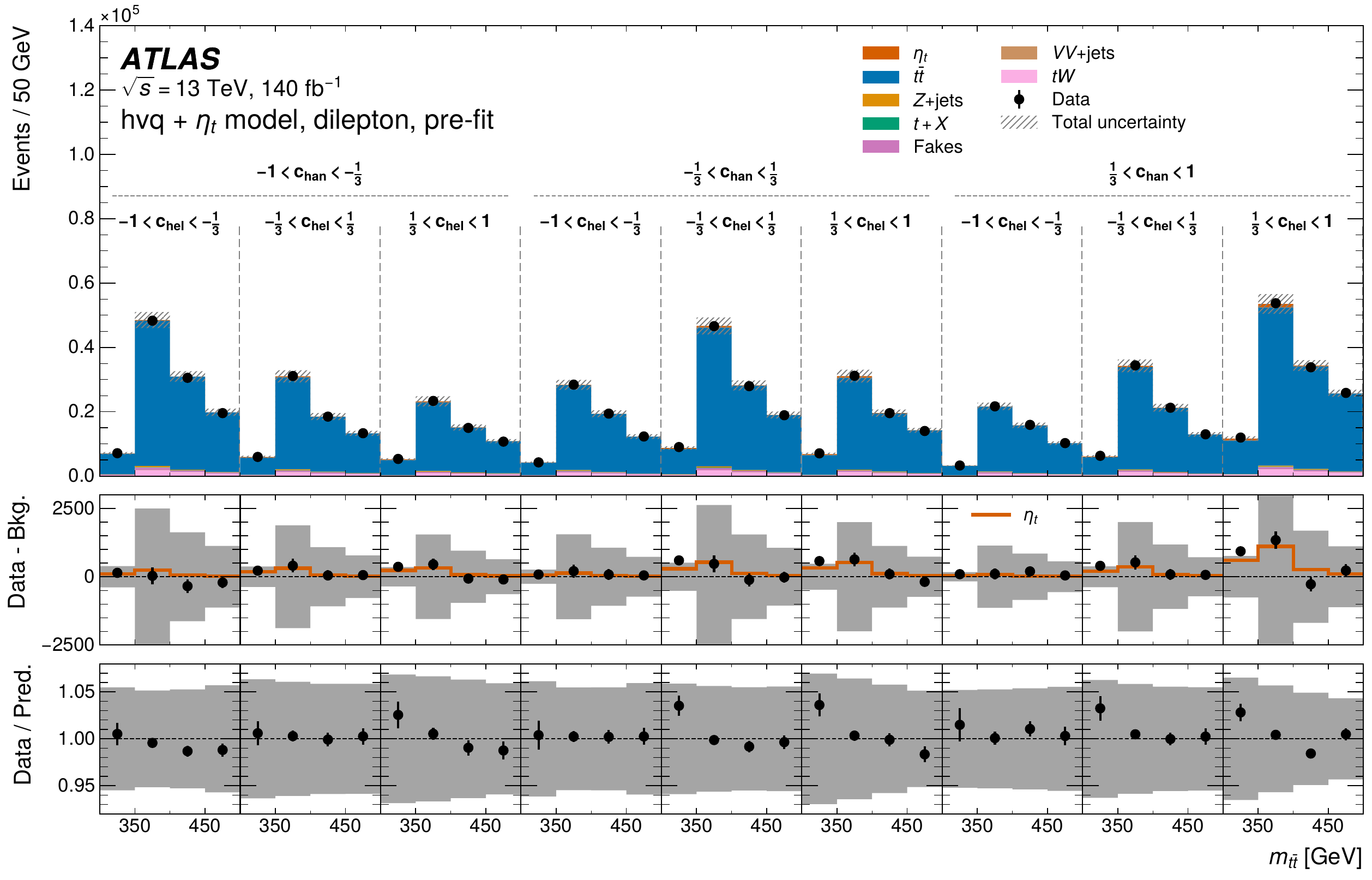}
\caption{Pre-fit distributions of \mttbar in the nine SRs (upper panel), together with a comparison between the \ttbar quasi-bound-state prediction, based on the simplified \toytoponium model, and the data, from which the \pQCD \ttbar contribution and background processes were subtracted (middle panel), and the ratio of the data and the extended \ttbar model including the \toytoponium contribution (lower panel). The error bars on the data markers represent the statistical uncertainty in the measurement, while the grey hashed and shaded bands represent the total systematic uncertainty in the prediction.}
\label{fig:toy:unblinded:hvq:summary2l_prefit}
\end{figure}

\begin{figure}[htbp]
\centering
\includegraphics[width=\textwidth]{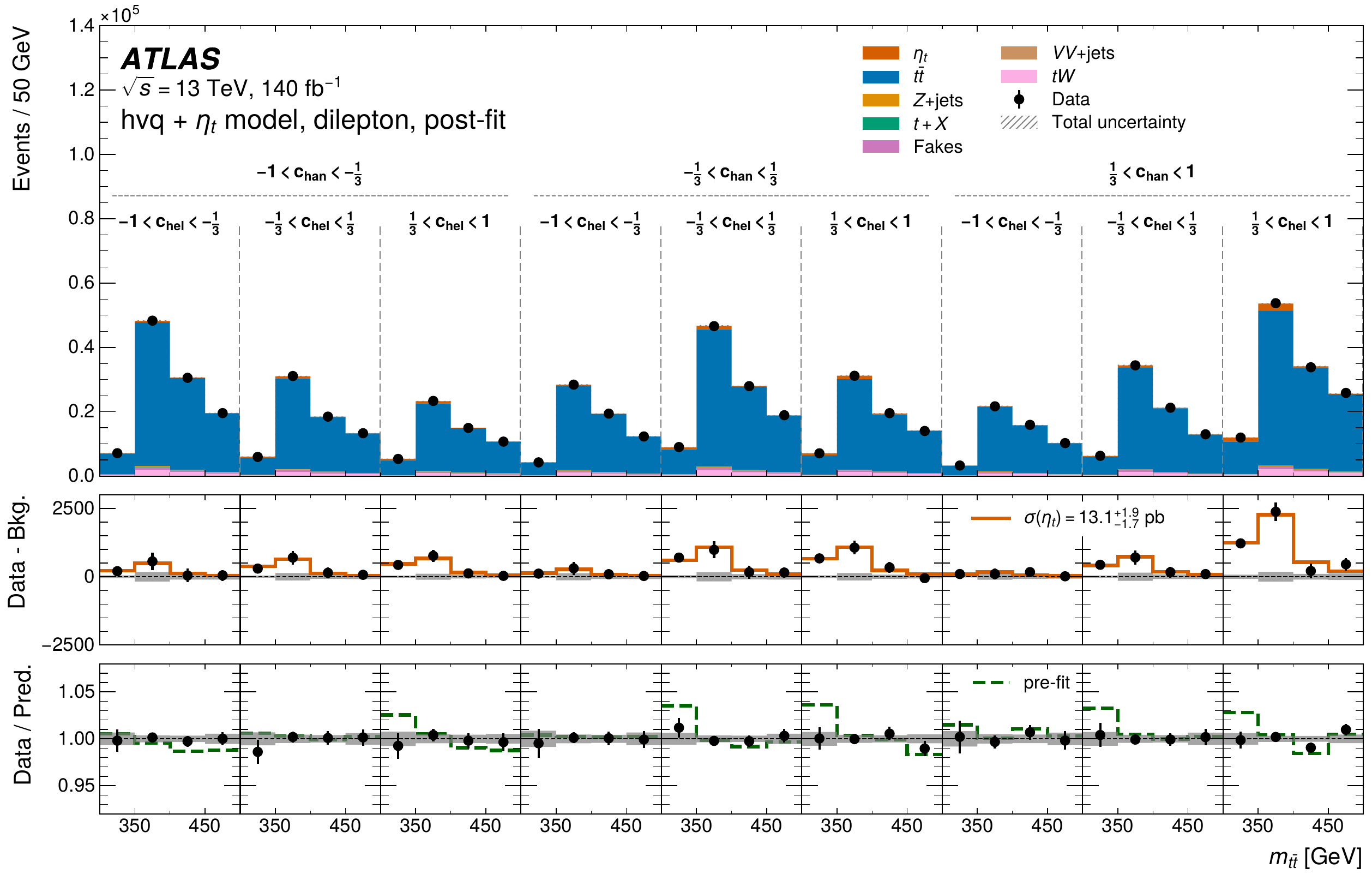}
\caption{
Post-fit distributions in the nine SRs (upper panel), together with a comparison between the \ttbar quasi-bound-state prediction, based on the simplified \toytoponium model, and the data, from which the \pQCD \ttbar contribution and background processes were subtracted (middle panel), and the ratio of the data to the extended \ttbar model including the \toytoponium contribution (lower panel). The error bars on the data markers represent the statistical uncertainty in the measurement, while the grey hashed and shaded bands represent the total systematic uncertainty in the prediction.}
\label{fig:toy:unblinded:hvq:summary2l_postfit}
\end{figure}

The systematic uncertainties with the largest impact on the extracted cross-section are shown in Figure~\ref{fig:toy:results:unblinded:hvq:ranking}.
Overall, the relative impacts and constraints are largely comparable with those obtained in the main fit, with the largest upward shift of approximately 0.5 pre-fit standard deviations observed for the NP on the uncorrelated part of the uncertainty in the modelling of the \ttbar decay and off-shell effects.
The impact of all uncertainties is summarised in Table~\ref{tab:toy:unblinded:grouped_uncs_hvq}.

\begin{figure}[htbp]
\centering
\includegraphics[width=0.8\textwidth]{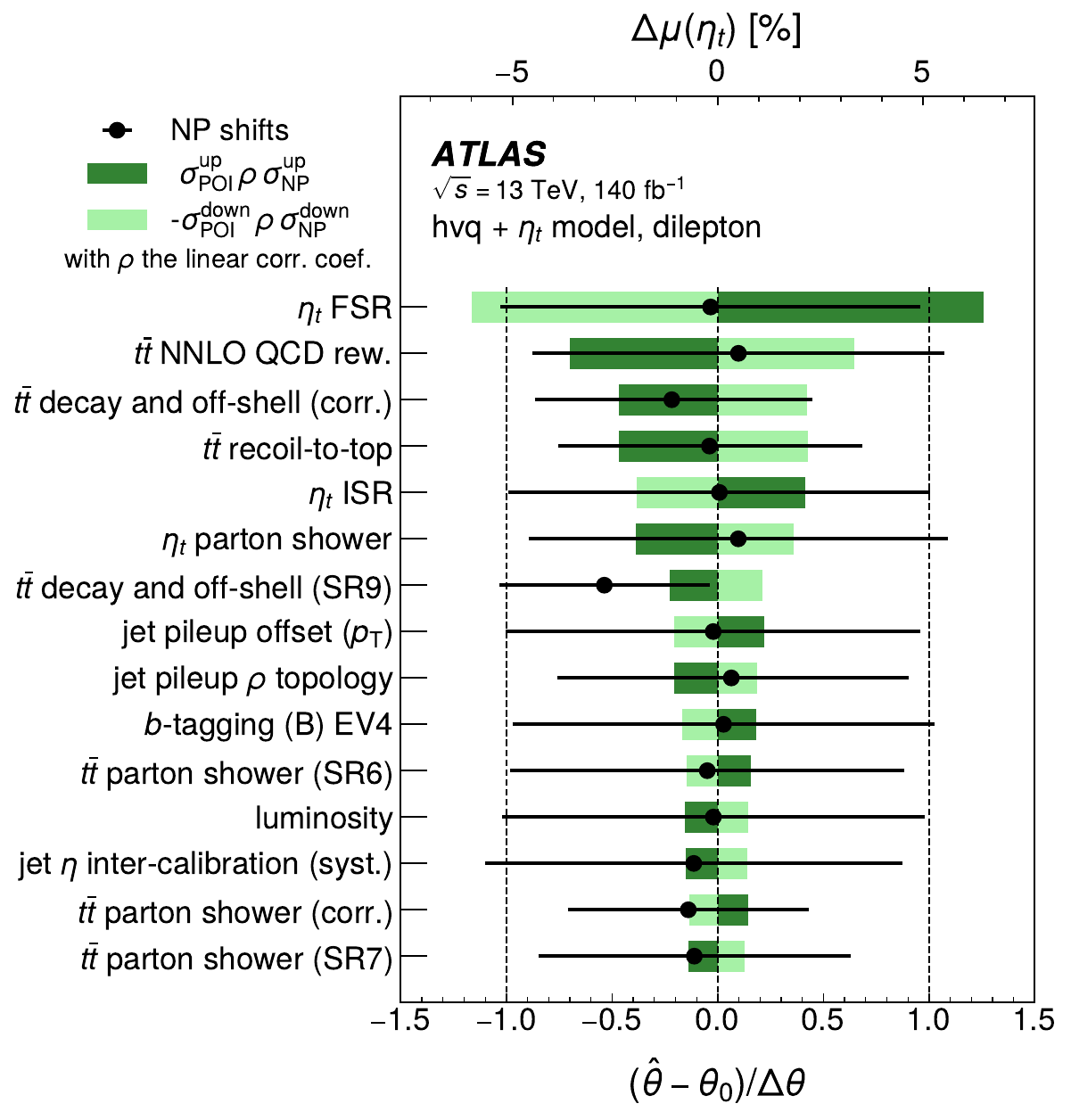}
\caption{Ranking of the 15 most impactful individual NPs on $\mu(\toytoponium)$ in the profile-likelihood fit when using the simplified model of \toytoponium.
The impact for a given NP is taken as the corresponding off-shell element of the fit covariance matrix divided by the pre-fit uncertainty~\cite{Pinto:2023yob} and given in percent on the upper scale. The black marker shows the shifts of the NPs relative to the nominal value $\theta_0$. They are shown together with their corresponding uncertainties on the lower scale. The SR numbers correspond to the bins introduced in Figure~\ref{fig:results:SRsplit}. For the flavour-tagging uncertainties, individual eigenvariations (EVs) are shown.}
\label{fig:toy:results:unblinded:hvq:ranking}
\end{figure}

\begin{table}[!htb]
\footnotesize
\caption{
Breakdown of the total uncertainty in the fitted normalisation factor $\mu (\toytoponium)$ into its statistical and systematic components.
The impact for a given NP is taken as the corresponding off-diagonal element of the fit covariance matrix divided by its pre-fit uncertainty.
The impact of a group of NPs is obtained by summing the impacts of all NPs in this category in quadrature.
The total systematic uncertainty is obtained by summing in quadrature the impacts of all NPs. The statistical uncertainty in the signal strength is calculated based
on the requirement that the sum in quadrature of the statistical uncertainty and the total systematic uncertainty must yield the total uncertainty in the signal strength as obtained from the initial profile likelihood fit in which all nuisance
parameters are allowed to float.
The uncertainties are symmetrised by taking the average of the upward and downward variations.
The category "Instrumental (other)" considers the uncertainty in the luminosity, pile-up reweighting and jet-vertex tagger efficiency.}
\label{tab:toy:unblinded:grouped_uncs_hvq}
\def\arraystretch{1.3}
\begin{center}
\begin{tabular}{lr}
\toprule
Category & Impact \\
\midrule
\toytoponium modelling & 6.7\% \\
\pQCD \ttbar modelling & 5.2\% \\
JES pile-up & 1.5\% \\
$b$-tagging & 1.3\% \\
Instrumental (other) & 0.9\% \\
JES $\eta$ inter-calibration & 0.8\% \\
Background normalisations & 0.6\% \\
JES flavour & 0.6\% \\
Leptons & 0.6\% \\
$tW$ modelling & 0.5\% \\
Jet energy scale (JES) & 0.5\% \\
Limited MC statistics & 0.3\% \\
Jet energy resolution (JER) & 0.1\% \\
\midrule
Total syst. uncertainties & 8.9\% \\
Total stat. uncertainties & 11\% \\
\bottomrule
\end{tabular}
\end{center}
\end{table}

\FloatBarrier


%

\FloatBarrier
\printbibliography
\clearpage
 
\begin{flushleft}
\hypersetup{urlcolor=black}
{\Large The ATLAS Collaboration}

\bigskip

\AtlasOrcid[0000-0002-6665-4934]{G.~Aad}$^\textrm{\scriptsize 103}$,
\AtlasOrcid[0000-0001-7616-1554]{E.~Aakvaag}$^\textrm{\scriptsize 17}$,
\AtlasOrcid[0000-0002-5888-2734]{B.~Abbott}$^\textrm{\scriptsize 123}$,
\AtlasOrcid[0000-0002-0287-5869]{S.~Abdelhameed}$^\textrm{\scriptsize 119a}$,
\AtlasOrcid[0000-0002-1002-1652]{K.~Abeling}$^\textrm{\scriptsize 55}$,
\AtlasOrcid[0000-0001-5763-2760]{N.J.~Abicht}$^\textrm{\scriptsize 49}$,
\AtlasOrcid[0000-0002-8496-9294]{S.H.~Abidi}$^\textrm{\scriptsize 30}$,
\AtlasOrcid[0009-0003-6578-220X]{M.~Aboelela}$^\textrm{\scriptsize 45}$,
\AtlasOrcid[0000-0002-9987-2292]{A.~Aboulhorma}$^\textrm{\scriptsize 36e}$,
\AtlasOrcid[0000-0001-5329-6640]{H.~Abramowicz}$^\textrm{\scriptsize 156}$
% ...
\bigskip
\\

% ...
$^{17}$Department for Physics and Technology, University of Bergen, Bergen; Norway.\\
% ...
$^{30}$Physics Department, Brookhaven National Laboratory, Upton NY; United States of America.\\
% ...
$^{36}$$^{(a)}$Facult\'e des Sciences Ain Chock, Universit\'e Hassan II de Casablanca;$^{(b)}$Facult\'{e} des Sciences, Universit\'{e} Ibn-Tofail, K\'{e}nitra;$^{(c)}$Facult\'e des Sciences Semlalia, Universit\'e Cadi Ayyad, LPHEA-Marrakech;$^{(d)}$LPMR, Facult\'e des Sciences, Universit\'e Mohamed Premier, Oujda;$^{(e)}$Facult\'e des sciences, Universit\'e Mohammed V, Rabat;$^{(f)}$Institute of Applied Physics, Mohammed VI Polytechnic University, Ben Guerir; Morocco.\\
% ...
$^{45}$Physics Department, Southern Methodist University, Dallas TX; United States of America.\\
% ...
$^{49}$Fakult\"{a}t Physik, Technische Universit{\"a}t Dortmund, Dortmund; Germany.\\
% ...
$^{55}$II. Physikalisches Institut, Georg-August-Universit\"{a}t G\"ottingen, G\"ottingen; Germany.\\
% ...
$^{103}$CPPM, Aix-Marseille Universit\'e, CNRS/IN2P3, Marseille; France.\\
% ...
$^{119}$$^{(a)}$New York University Abu Dhabi, Abu Dhabi;$^{(b)}$United Arab Emirates University, Al Ain; United Arab Emirates.\\
% ...
$^{123}$Homer L. Dodge Department of Physics and Astronomy, University of Oklahoma, Norman OK; United States of America.\\
% ...
$^{156}$Raymond and Beverly Sackler School of Physics and Astronomy, Tel Aviv University, Tel Aviv; Israel.\\
% ...
$^{a}$ Also at Affiliated with an institute formerly covered by a cooperation agreement with CERN.\\
$^{b}$ Also at An-Najah National University, Nablus; Palestine.\\
$^{c}$ Also at Borough of Manhattan Community College, City University of New York, New York NY; United States of America.\\
$^{d}$ Also at Center for Interdisciplinary Research and Innovation (CIRI-AUTH), Thessaloniki; Greece.\\
$^{e}$ Also at Centre of Physics of the Universities of Minho and Porto (CF-UM-UP); Portugal.\\
$^{f}$ Also at CERN, Geneva; Switzerland.\\
% ...

\end{flushleft}


%

%
%

%

%
%
%
%

\end{document}